\documentclass[published]{JHEP3} 

\JHEP{00(2008)000}

\JHEPspecialurl{http://jhep.sissa.it/JOURNAL/JHEP3.tar.gz}

\usepackage{epsfig,multicol,bbm}

\newcommand\fverb{\setbox\fverbbox=\hbox\bgroup\verb}
\newcommand\fverbdo{\egroup\medskip\noindent%
			\fbox{\unhbox\fverbbox}\ }
\newcommand\fverbit{\egroup\item[\fbox{\unhbox\fverbbox}]}
\newbox\fverbbox

\newcommand{\st}[2]{\stackrel{\mbox{\tiny (#1)}}{#2}\hspace{-0.1cm}}
\def\h#1{\hbox{${}^{#1}$H}}

\def\h502{\hbox{$ h^{2}_{50}$}}

\def\la{\mathrel{\mathpalette\fun <}}
\def\ga{\mathrel{\mathpalette\fun >}}
\def\fun#1#2{\lower3.6pt\vbox{\baselineskip0pt\lineskip.9pt
  \ialign{$\mathsurround=0pt#1\hfil##\hfil$\crcr#2\crcr\sim\crcr}}}

\newcommand{\adota}{\frac{\dot a}{a}}
\newcommand{\vct}[1]{\mbox{\boldmath${#1}$}}

\title{%
Neutrino Masses from Cosmological Probes \\
in Interacting Neutrino Dark-Energy Models
}%

\author{Kiyotomo Ichiki \\
Research Center for the Early Universe, University of Tokyo, 7-3-1
Hongo, Bunkyo-ku, Tokyo 113-0033, Japan \\
E-mail: \email{ichiki@resceu.s.u-tokyo.ac.jp} }
\author{ Yong-Yeon Keum \\
Division of Theoretical Astronomy,
National Astronomical Observatory of Japan, Mitaka, Tokyo 181-8588, Japan \\
Department of Physics, National Taiwan University, Taipei, Taiwan 10672, ROC \\
E-mail: \email{yykeum@phys.ntu.edu.tw (Corresponding Author)} }

\received{March 18, 2008} 		
\accepted{June 04, 2008}		

\preprint{\hepth{9912999}}	

\abstract{
We investigate whether interaction between massive neutrinos and quintessence
scalar field is the origin of the late time accelerated expansion of the universe.
We present explicit formulas of the cosmological linear perturbation
theory in the neutrinos probes of dark-energy model, and calculate
cosmic microwave background anisotropies and matter power spectra.
In these models, the evolution of the mass of neutrinos is
determined by the quintessence scalar field, which is responsible
for a varying effective equation of states; $\omega_{eff}(z)$ goes down lesser than -1. 
We consider several types of
scalar field potential and put constraints on the coupling
parameter between neutrinos and dark energy.
By combining data from cosmic microwave background (CMB) experiments including
the WMAP 3-year results, large scale structure with 2dFGRS data sets,
we constrain the hypothesis of massive neutrinos in the mass-varying neutrino
scenario.
Assuming the flatness
of the universe, the constraint we can derive from the current
observation is $\sum m_{\nu} < 0.45$ eV at 1$\sigma$ ($0.87$ eV at 2$\sigma$) confidence level for the sum over three species of neutrinos. 
The dynamics of scalar field and the impact of scalar field perturbations on cosmic 
microwave background anisotropies are discussed. We also discuss on the instability
issue and confirm that neutrinos are stable against 
the density fluctuation in our model.
}


\begin{document} 


\newpage

\section{Introduction}
After Type Ia Super Novae (SNIa) \cite{sn1a} and Cosmic Microwave
Background \cite{wmap} 
observations in the last decade, the discovery of an accelerating
expansion of the universe is a major challenge to particle physics
and cosmology. Many models to explain such an accelerating expansion
have been proposed so far, and they are mainly categorized into three: 
namely,  a non-zero  cosmological constant \cite{lambda}, a dynamical
cosmological constant (quintessence scalar field) \cite{{quintessence},{ratra-peebles:1988}}, 
modifications of Einstein Theory of Gravity \cite{mgrav}. 
One often call
what drives the late-time cosmic acceleration as {\it dark energy}.

While the existence of such dark energy component has become observationally
evident, the current observational data sets are consistent with all of
the three possibilities above. The first observational goal to be achieved
is therefore to know whether the dark energy is
cosmological constant or dynamical component; 
in other words, the equation of state parameter of dark energy,
$w=P/\rho$, is $-1$ or not. The scalar field model like quintessence
is a simple model with time dependent $w$, which is generally larger
than $-1$. Because the different $w$ leads to a different expansion
history of the universe, the geometrical measurements of  
cosmic expansion through observations of SNIa, CMB, and Baryon Acoustic
Oscillations (BAO) can give us tight
constraints on $w$. Recent compilations of those data sets suggest
that the dark energy is consistent with cosmological constant
\cite{seljak:0604335,Percival:2007yw}. 

Further, if the dark energy is dynamical component
like a scalar field, it should carry its density
fluctuations. Thus, the probes of density fluctuations near
the present epoch, such as cross correlation studies of the integrated
Sachs-Wolfe effect \cite{Crittenden:1995ak,Hu:2004yd} and the power of
Large Scale Structure (LSS) \cite{Takada:2006xs}, can also provide useful
information to discriminate between cosmological constant and others. 
Yet, current observational data can give only poor constraints on   
the properties of dark energy fluctuations \cite{Hannestad:2005ak,Ichiki:2007vn}. 

Another interesting way to study the scalar field dark energy models is
to investigate the coupling between the dark energy and the other matter
fields. In fact, a number of models which realize the interaction
between dark energy and dark matter, or even visible matters, have been
proposed so far
\cite{Carroll:1998zi,Bean:2000zm,Farrar:2003uw,das-khoury:2006,Lee:2006za}. 
Observations of the effects of these interactions will offer an unique
opportunity to detect a cosmological scalar field
\cite{Carroll:1998zi,Liu:2006uh}. 

An interesting model for the interacting dark-energy with massive neutrinos 
was proposed by Fardon, Kaplan, Nelson and Weiner \cite{Fardon:2003eh,mavanu}, 
in which neutrinos are an integral component of the dark energy.
They consider the existence of new particles (accelerons with masses of about
$10^{-3}$  eV) that generate the dark energy, and the possible interactions between
these particles and the known particles of the Standard Model.
As far as quarks and charged leptons are concern, there are severe constraints on the 
strength of these interactions. However, neutrinos-weakly interacting very-low-mass particles- 
are exceptional, and cosmologically relevant couplings are still possible for them.
This type has the scalar field adiabatically tracking the minimum of the effective potential, 
and, however, finally runs the adiabatic instability when the mass-varying neutrinos 
become non-relativistic, since the mass of scalar field must be much larger than
the Hubble expansion rate for the adiabatic approximation to be valid 
\cite{Afshordi:2005ym,Bean:2007ny}.

The other type has the dark-energy provided by a quintessence scalar field evolving in 
a nearly flat potential, whose derivatives must satisfy the slow-roll condition.
In our analysis, we adopt the latter type, in which the potential is somehow 
finely tuned so that the rate of scalar field evolution comes out 
to be comparable to the Hubble rate. Naturally couplings between the quintessence field
to neutrinos leads to a back reaction on the potential, which was not taken into account 
in previous works \cite{Brookfield-b, Zhao:2006zf} with slow-rolling quintessence field, 
but we take into account fully the neutrino contribution to the effective potential
in this work. We calculate explicitly Cosmic Microwave Background (CMB) radiation
and Large Scale Structures (LSS) within cosmological perturbation theory.
We first analyze constraints on a set of models, which has not been correctly done before,
with cosmological observation data with WMAP3-year data and 2dFGRS data sets.

In this paper, after reviewing shortly the main idea of the three
possible candidates of dark energy and their cosmological phenomena in
section II, we discuss the interacting dark energy model, paying
particular attention to the interacting mechanism between
dark-energy with a hot dark-matter (neutrinos) in section III. The
evolution of the mass of neutrinos is determined by the quintessence
scalar filed, which is responsible for a varying equation of states: 
$\omega_{eff}(z)$ goes down lesser than -1. 
Recently, perturbation equations for this class of models are
nicely presented by Brookfield et al. \cite{Brookfield-b}, (see also
\cite{Zhao:2006zf}) which are necessary to compute CMB and LSS
spectra. A main difference here from their works is that we correctly
take into account the scattering term in the geodesic equation of
neutrinos, which was omitted there (see, however, the Erratum of Ref. \cite{Brookfield:2005bz}).
We will show and remark  in section IV that this leads significant
differences in the resultant spectra and hence the different
observational constraints. In section V, we discuss three different
types of quintessence potential, namely, an inverse power law potential,
a supergravity potential, and an exponential type potential. By
computing CMB and LSS spectra with these quintessential potentials
and comparing them to the latest observations, the constraints on the
present mass of neutrinos and coupling parameters are derived. We also
show that the equation 
of state $w_{z=0}$ can reach down to $-1$, which is consistent with
experimental observations. In section VI, we discuss on the neutrino
mass bound in our interacting dark-energy models.
In appendix A, we show explicit derivation of geodesic
equation in the Boltzmann equation which contains a new contribution from
the time-dependent neutrino mass
and some formulas of the varying neutrino mass in early stage of universe
in Appendix B. Finally 
the explicit calculation for
the consistency check of our calculations in section III is shown
in Appendix C.

\section{Three possible solutions for Accelerating Universe:}
Recent observations of SNIa and CMB radiation have provided strong
evidence that we live now in an accelerating 
and almost flat universe. In general, one believes that the
dominance of a dark-energy component with negative pressure in the
present era is responsible for the universe's accelerated expansion.

There are mainly three possible solutions to explain the accelerating
universe. The Einstein Equation in General Relativity is given by the
following form: 
\begin{equation}
G_{\mu\nu} \,\,= \,\, R_{\mu\nu} -{1 \over 2} R \, g_{\mu\nu} \hspace{2mm}
= \hspace{2mm} 8 \pi G \, T_{\mu\nu} \hspace{2mm} + \hspace{2mm} \Lambda \, g_{\mu\nu},
\label{eq:einstein}
\end{equation}
Here, $G_{\mu\nu}$ term contains the information of geometrical
structure, the energy-momentum tensor $T_{\mu\nu}$ keeps the
information of matter distributions, and the last term is so called
the cosmological constant which contain the information of non-zero
vacuum energy. By writing the Einstein equation with a flat
Robertson-Walker metric, one can drive a simple relation:
\begin{equation}
{\ddot{R} \over R} = -{4 \pi G \over 3} (\rho +3p) +{\Lambda \over
3}.
\end{equation}
In order to get the accelerating expansion, either positive cosmological 
constant $\Lambda$ ($\omega_{\Lambda} = P/\rho=-1$) or a new concept of
dark-energy with the negative pressure 
($\omega_{\phi} <-1/3$) needs to be introduced. Another solution can
be given by the modification of geometrical structure which can
provide a repulsive source of gravitational force.  In this case,
the attractive gravitational force term is dominant in early stage
of universe, however at later time near the present era, repulsive
term become important at cosmological scales and drives universe to be
expanded with an acceleration \cite{Dvali:2000hr}.
Also we can consider extra-energy density contributions from bulk space
in so-called 'cosmological brane world' models, which 
can modify the Friedman equation as $H^2 \propto \rho + \rho^{'}$
\cite{Brane-World,MOND}.

In summary, we have three different solutions for the accelerating
expansion of our universe as mentioned in the introduction.
 Probing for the origin of accelerating universe is the most
important and challenged problem in high energy physics and
cosmology at present. The detailed explanation and many references are in
a nice review on dark energy \cite{review-DE}.  
In this paper, we concentrate on the second class of solutions using the 
quintessence field. In the present epoch, the potential term becomes
important than the kinetic term, which can easily explain the negative
pressure with $\omega_{\phi}^0 \simeq -1$. However there are many
different versions of quintessence field:
K-essence\cite{COY,k-essence}, phantom\cite{phantom},
quintom\cite{quintom}, and so on, and justifying the origin of
dark-energy from experimental observations is really a difficult
job. 

\section{Cosmological perturbations in Interacting Dark-Energy with Neutrinos:}
As explained in previous section, it is really difficult to probe
the origin of dark-energy when the dark-energy doesn't interact with
other matters at all. Here we investigate the cosmological
implication of an idea of the dark-energy interacting with neutrinos
\cite{{Fardon:2003eh},{mavanu}}. For simplicity, we consider the
case that dark-energy and neutrinos are coupled such that the mass
of the neutrinos is a function of the scalar field which drives the
late time accelerated expansion of the universe. In previous works
by Fardon et al. \cite{Fardon:2003eh} and R. Peccei \cite{mavanu},
the kinetic energy term was ignored and potential term was treated as a
dynamical cosmological constant, which can be applicable for the 
dynamics near present epoch. However the kinetic contributions
become important to describe cosmological perturbations in early
stage of universe, which is fully considered in our analysis.

\subsection{Background Equations}
Equations for quintessence scalar field are given by
\begin{eqnarray}
\ddot{\phi}&+&2{\cal H}\dot\phi+a^2\frac{d V_{\rm eff}(\phi)}{d\phi}=0~,
 \label{eq:Qddot}\\
V_{\rm eff}(\phi)&=&V(\phi)+V_{\rm I}(\phi)~,\\
V_{\rm I}(\phi)&=&a^{-4}\int\frac{d^3q}{(2\pi)^3}\sqrt{q^2+a^2
 m_\nu^2(\phi)}f(q)~,\\
m_\nu(\phi) &=& \bar m_i e^{\beta\frac{\phi}{M_{\rm pl}}}~,
\end{eqnarray}
where $V(\phi)$ is the potential of quintessence scalar field, $V_{\rm
I}(\phi)$ is additional potential due to the coupling to neutrino
particles \cite{Fardon:2003eh,Bi:2003yr},
and $m_\nu(\phi)$ is the mass of neutrino coupled to the scalar field,
where we assume the exponential coupling with a coupling parameter
$\beta$. 
${\cal H}$ is $\frac{\dot a}{a}$, where the dot represents the
derivative with respect to the conformal time $\tau$.

Energy densities of mass varying neutrinos  and quintessence scalar
field are described as
\begin{eqnarray}
\rho_\nu &=& a^{-4}\int \frac{d^3 q}{(2\pi)^3} \sqrt{q^2+a^2m_\nu^2} f_0(q)~, \label{eq:rho_nu}\\
3P_\nu &=& a^{-4}\int \frac{d^3 q}{(2\pi)^3} \frac{q^2}{\sqrt{q^2+a^2m_\nu^2}}
 f_0(q)~, \label{eq:P_nu}\\
\rho_\phi &=& \frac{1}{2a^2}\dot\phi^2+V(\phi)~,\label{eq:rho_phi}\\
P_\phi &=& \frac{1}{2a^2}\dot\phi^2-V(\phi)~.
\end{eqnarray}
From equations (\ref{eq:rho_nu}) and (\ref{eq:P_nu}), the equation of
motion for the background energy density of neutrinos is given by
\begin{equation}
\dot\rho_{\nu}+3{\cal H}(\rho_\nu+P_\nu)=\frac{\partial \ln
 m_\nu}{\partial \phi}\dot\phi(\rho_\nu -3P_\nu)~.
\end{equation}
\subsection{Perturbation equations:}
\subsubsection{perturbations in the metric} We work in the
synchronous gauge and line element is
\begin{equation}
 ds^2 = a^2(\tau)\left[-d\tau^2 + (\delta_{ij}+h_{ij})dx^i dx^j\right]~,
\end{equation}
In this metric the Christoffel symbols which have non-zero values are
\begin{eqnarray}
\Gamma^0_{00}&=&\adota~, \\
\Gamma^0_{ij}&=&\adota \delta_{ij}+\frac{\dot
 a}{a}h_{ij}+\frac{1}{2}\dot h_{ij}~, \\
\Gamma^i_{0j}&=&\adota\delta^i_j+\frac{1}{2}\dot h_{ij}~,\\
\Gamma^i_{jk}&=&\frac{1}{2}\delta^{ia}(h_{ka,j}+h_{aj,k}-h_{jk,a})~,
\end{eqnarray}
where dot denotes conformal time derivative.
For CMB anisotropies we mainly consider the scalar type perturbations. We
introduce two scalar fields, $h(\vct k, \tau)$ and $\eta(\vct k, \tau)$, in
k-space and write the scalar mode of $h_{ij}$ as a Fourier integral
\cite{Ma:1995ey}
\begin{equation}
h_{ij}(\vct x,\tau)=\int d^3k e^{i\vct{k}\cdot\vct{x}}\left[\vct{\hat
 k}_i\vct{\hat k}_j h(\vct k,\tau)+(\vct{\hat k}_i \vct{\hat k}_j
 -\frac{1}{3}\delta_{ij})6\eta(\vct{k},\tau) \right]~,
\end{equation}
where $\vct{k} =k\vct{\hat k}$ with $\hat{k}^i \hat{k}_i =1$.
\subsubsection{perturbations in quintessence} 
The equation of quintessence scalar field is given by
\begin{equation}
 \Box \phi - V_{\rm eff}(\phi)=0~.
\end{equation}
Let us write the scalar field as a sum of background value and
perturbations around it, $\phi(\vct x,\tau)=\phi(\tau)+\delta \phi(\vct
x,\tau)$.
The perturbation equation is then described as
\begin{equation}
\frac{1}{a^2}\ddot{\delta \phi}+\frac{2}{a^2}{\cal H}\dot\delta\phi -
 \frac{1}{a^2}\nabla^2(\delta\phi)+\frac{1}{2a^2}\dot{h}\dot\phi+\frac{d^2
 V}{d\phi^2}\delta\phi +\delta\left(\frac{dV_{\rm I}}{d\phi}\right)=0~ \label{eq:dQddot},
\end{equation}
where
\begin{eqnarray}
\frac{d V_{\rm I}}{d\phi} &=& a^{-4}\int
 \frac{d^3q}{(2\pi)^3}\frac{\partial \epsilon(q,\phi)}{\partial \phi}
 f(q)~,\label{eq:dvidphi}\\
\epsilon(q,\phi) &=& \sqrt{q^2+a^2 m_\nu^2(\phi)} ~,\\
\frac{\partial \epsilon(q,\phi)}{\partial \phi} &=& \frac{a^2
 m^2_\nu(\phi)}{\epsilon(q,\phi)} \frac{\partial \ln m_\nu}{\partial \phi}~.
\end{eqnarray}
To describe $\delta\left(\frac{dV_{\rm I}}{d\phi}\right)$,
we shall write the distribution function of neutrinos with background
distribution and perturbation around it as
\begin{equation}
f(x^i, \tau, q, n_j )= f_0(\tau,q)(1+\Psi(x^i, \tau, q, n_j))~.
\end{equation}
Then we can write
\begin{equation}
\delta\left(\frac{dV_{\rm I}}{d\phi}\right)
 =a^{-4}\int\frac{d^3q}{(2\pi)^3}
 \frac{\partial^2 \epsilon}{\partial \phi^2}\delta\phi f_0
 + a^{-4}\int\frac{d^3q}{(2\pi)^3}\frac{\partial \epsilon}{\partial
 \phi}f_0\Psi~, \label{eq:delta(dvidphi)}
\end{equation}
where
\begin{eqnarray}
\frac{\partial^2 \epsilon}{\partial \phi^2}&=&\frac{a^2}{\epsilon}\left(\frac{\partial m_\nu}{\partial \phi}\right)^2 +\frac{a^2 m_\nu}{\epsilon}\left(\frac{\partial^2 m_\nu}{\partial \phi^2}\right)
-\frac{a^2 m_\nu}{\epsilon^2}\left(\frac{\partial \epsilon}{\partial \phi}\right)\left(\frac{\partial m_\nu}{\partial \phi}\right)~.
\end{eqnarray}
For numerical purpose it is useful to rewrite the equations
(\ref{eq:dvidphi}) and (\ref{eq:delta(dvidphi)}) as
\begin{eqnarray}
\frac{dV_{\rm I}}{d\phi} &=& \frac{\partial \ln m_\nu}{\partial
 \phi}(\rho_\nu -3P_\nu)~, \\
\delta\left(\frac{dV_{\rm I}}{d\phi}\right) &=&
\frac{\partial^2 \ln m_\nu}{\partial
 \phi^2} \delta\phi (\rho_\nu-3P_\nu)
+ \frac{\partial \ln m_\nu}{\partial
 \phi}(\delta\rho_\nu-3\delta P_\nu)~.
\end{eqnarray}
Note that perturbation fluid variables in mass varying neutrinos are given by
\begin{eqnarray}
\delta\rho_\nu&=&a^{-4}\int\frac{d^3 q}{(2\pi)^3} \epsilon
 f_0(q)\Psi+a^{-4}\int\frac{d^3 q}{(2\pi)^3}\frac{\partial
 \epsilon}{\partial \phi}\delta\phi f_0 ~, \label{eq:delta_rho_nu}\\
3\delta P_\nu&=&a^{-4}\int\frac{d^3 q}{(2\pi)^3} \frac{q^2}{\epsilon}
 f_0(q)\Psi-a^{-4}\int\frac{d^3
 q}{(2\pi)^3}\frac{q^2}{\epsilon^2}\frac{\partial \epsilon}{\partial
 \phi}\delta\phi f_0 ~.
\end{eqnarray}
The energy momentum tensor of quintessence is given by
\begin{equation}
 T^\mu_\nu = g^{\mu\alpha}\phi_{,\alpha}\phi_{,\nu}-\frac{1}{2}\left(\phi^{,\alpha}\phi_{,\alpha}+2V(\phi)\right)\delta^\mu_\nu~,
\end{equation}
and its perturbation is
\begin{eqnarray}
 \delta T^\mu_\nu&=&g_{(0)}^{\mu \alpha}\delta\phi_{,\alpha}\phi_{,\nu}
  +g_{(0)}^{\mu \alpha}\phi_{,\alpha}\delta \phi_{,\nu}
  +\delta g^{\mu \alpha}\phi_{,\alpha}\phi_{,\nu} \nonumber \\
  & & \hspace{5mm} -\frac{1}{2}\left(\delta\phi^{,\alpha}\phi_{,\alpha}
          +\phi^{,\alpha}\delta\phi_{,\alpha}
          +2\frac{dV}{d\phi}\delta\phi\right)\delta^\mu_\nu~.
\end{eqnarray}
This gives perturbations of quintessence in fluid variables as
\begin{eqnarray}
 \delta\rho_\phi&=&-\delta T^0_0 =
 \frac{1}{a^2}\dot\phi\dot{\delta\phi}+\frac{dV}{d\phi}\delta\phi~, \\
 \delta P_\phi&=&-\delta T^0_0/3 =
 \frac{1}{a^2}\dot\phi\dot{\delta\phi}-\frac{dV}{d\phi}\delta\phi~, \\
 (\rho_\phi +P_\phi)\theta_\phi&=&ik^i\delta
 T^0_i=\frac{k^2}{a^2}\dot\phi\delta\phi~,\\
 \Sigma^i_j&=&T^i_j-\delta^i_j T^k_k/3=0~.
\end{eqnarray}

\subsection{Boltzmann Equation for Mass Varying Neutrino}
One has to consider Boltzmann equation to solve the evolution of
Mass Varying Neutrinos. A distribution function is written in terms of time ($\tau$),
positions ($x^i$) and their conjugate momentum ($P_i$). The
conjugate momentum is defined as spatial parts of the 4-momentum
with lower indices, i.e., $P_i = mU_i$, where
$U_i=dx_i/(-ds^2)^{1/2}$. We also introduce locally orthonormal
coordinate $X^\mu = (t,r^i)$, and we write the energy and the
momentum in this coordinate as $(E,p^i)$, where
$E=\sqrt{p^2+m_\nu^2}$. The relations of these variables in
synchronous gauge are given by \cite{Ma:1995ey},
\begin{eqnarray}
 P_0 &=& -aE~, \\
 P_i &=& a(\delta_{ij}+\frac{1}{2}h_{ij})p^{j}~.
\end{eqnarray}
Next we define comoving energy and momentum $(\epsilon, q_i)$ as
\begin{eqnarray}
 \epsilon &=& aE = \sqrt{q^2+a^2 m_\nu^2}~,\\
 q_i &=& ap_i~.
\end{eqnarray}
Hereafter, we shall use ($x^i,q,n_j,\tau$) as phase space variables,
replacing $f(x^i,P_j,\tau)$ by $f(x^i,q,n_j,\tau)$. Here
we have splitted the comoving momentum $q_j$ into its magnitude and
direction: $q_j = qn_j$, where $n^i n_i=1$.
The Boltzmann equation is
\begin{equation}
 \frac{Df}{D\tau}=\frac{\partial f}{\partial
 \tau}+ \frac{dx^i}{d\tau}\frac{\partial f}{\partial
 x^i}
+\frac{dq}{d\tau}\frac{\partial f}{\partial
 q}+\frac{dn_i}{d\tau}\frac{\partial f}{\partial
 n_i}=\left(\frac{\partial f}{\partial \tau}\right)_C~.
\end{equation}
in terms of these variables.
From the time component of geodesic equation \cite{Anderson:1997un} (see
also Appendix A),
\begin{equation}
 \frac{1}{2}\frac{d}{d\tau}\left(P^0\right)^2=-\Gamma^0_{\alpha\beta}P^\alpha
  P^\beta - m g^{0 \mu}m_{,\mu}~,
\label{eq:eq-a}
\end{equation}
and the relation $P^0=a^{-2}\epsilon=a^{-2}\sqrt{q^2+a^2 m_\nu^2}$, we
have
\begin{equation}
 \frac{dq}{d\tau}=-\frac{1}{2}\dot{h_{ij}}qn^in^j-a^2
  \frac{m}{q}\frac{\partial m}{\partial x^i}\frac{dx^i}{d \tau} ~.
\label{eq:eq-b}
\end{equation}
Our analytic formulas in eqs.(\ref{eq:eq-a}-\ref{eq:eq-b}) are different
from those of \cite{Brookfield-b} and \cite{Zhao:2006zf}, since they
have omitted the contribution of the varying neutrino mass term. We
shall show later this term also give an important contribution in the
first order perturbation of the Boltzmann equation. 

We will write down each term up to ${\cal O}(h)$:
\begin{eqnarray}
 \frac{\partial f}{\partial \tau} &=&\frac{\partial f_0}{\partial
 \tau}+f_0 \frac{\partial \Psi}{\partial \tau} +\frac{\partial
 f_0}{\partial \tau}\Psi~, \nonumber \\
 \frac{dx^i}{d\tau}\frac{\partial f}{\partial
 x^i}&=&\frac{q}{\epsilon}n^i \times f_0\frac{\partial \Psi}{\partial
 x^i} ~,\nonumber \\
 \frac{dq}{d\tau}\frac{\partial f}{\partial
 q}&=&\left(-a^2\frac{m_\nu}{q}\frac{\partial m_\nu}{\partial x^i}\frac{dx^i}{d\tau}-\frac{1}{2}\dot{h_{ij}}qn^in^j\right)
\times\frac{\partial f_0}{\partial q}~, \nonumber \\
 \frac{dn_i}{d\tau}\frac{\partial f}{\partial
 n_i}&=& {\cal O}(h^2)~.
\end{eqnarray}
We note that $\frac{\partial f}{\partial
 x^i}$ and $\frac{dq}{d\tau}$ are ${\cal O}(h)$.

\subsubsection{Background equations} From the equations above, the
zeroth-order Boltzmann equation is
\begin{equation}
 \frac{\partial f_0}{\partial\tau}=0~.
\label{eq:Boltz0}
\end{equation}
The Fermi-Dirac distribution
\begin{equation}
 f_0=f_0(\epsilon)=\frac{g_s}{h_{\rm P}^3}\frac{1}{e^{\epsilon/k_{\rm B}T_0}+1}~,
\end{equation}
can be a solution.
Here $g_s$ is the number of spin degrees of freedom, $h_{\rm P}$ and
$k_{\rm B}$ are the Planck and the Boltzmann constants.

\subsubsection{Perturbation Equations} 
The first-order Boltzmann equation is
\begin{eqnarray}
 \frac{\partial \Psi}{\partial
 \tau}&+&i\frac{q}{\epsilon}(\vct{\hat{n}}\cdot\vct{k})\Psi+\left(\dot\eta-(\vct{\hat
 k}\cdot\vct{\hat n})^2\frac{\dot h+6\dot\eta}{2}\right)\frac{\partial \ln
 f_0}{\partial \ln q} \nonumber \\
&& \hspace{5mm} -i\frac{q}{\epsilon}(\vct{\hat{n}}\cdot\vct{k})k\delta\phi\frac{a^2
 m^2}{q^2}\frac{\partial \ln m}{\partial \phi}\frac{\partial \ln
 f_0}{\partial \ln q} = 0~.
\label{eq:boltzmann}
\end{eqnarray}
Following previous studies, we shall assume that the initial momentum
dependence is axially symmetric so that $\Psi$ depends on
$\vct{q}=q\vct{\hat n}$ only through $q$ and $\vct{\hat k}\cdot\vct{\hat
n}$.
With this assumption, we expand the perturbation of distribution
function, $\Psi$, in a Legendre series,
\begin{equation}
 \Psi(\vct{k},\vct{\hat n},q,\tau)=\sum (-i)^\ell(2\ell+1)\Psi_\ell(\vct{k},q,\tau)P_\ell(\vct{\hat{k}}\cdot\vct{\hat{n}})~.
\end{equation}
Then we obtain the hierarchy for Mass Varying Neutrinos
\begin{eqnarray}
 \dot{\Psi_0}&=&-\frac{q}{\epsilon}k\Psi_1
                +\frac{\dot h}{6}\frac{\partial \ln{f_0}}{\partial\ln{q}}~, \label{eq:dot_Psi_0}\\
\dot{\Psi_1}&=&\frac{1}{3}\frac{q}{\epsilon}k\left(\Psi_0-2\Psi_2\right)
              + \kappa~, \label{eq:dot_Psi_1}\\
\dot{\Psi_2}&=&\frac{1}{5}\frac{q}{\epsilon}k(2\Psi_1-3\Psi_3)-\left(\frac{1}{15}\dot{h}+\frac{2}{5}\dot{\eta}\right)\frac{\partial \ln{f_0}}{\partial \ln{q}}~,\\
\dot{\Psi_\ell}&=&\frac{q}{\epsilon}k\left(\frac{\ell}{2\ell+1}\Psi_{\ell-1}-\frac{\ell+1}{2\ell+1}\Psi_{\ell+1}\right)~.
\end{eqnarray}
where
\begin{equation}
\kappa = -\frac{1}{3}\frac{q}{\epsilon}k\frac{a^2 m^2}{q^2}\delta\phi
 \frac{\partial \ln m_\nu}{\partial \phi}\frac{\partial \ln
 f_0}{\partial \ln q}\label{eq:kappa}~.
\end{equation}
Here we used the recursion relation
\begin{equation}
 (\ell+1)P_{\ell+1}(\mu)=(2\ell+1)\mu P_\ell(\mu)-\ell P_{\ell-1}(\mu)~.
\end{equation}
We have to solve these equations with a $q$-grid for every wave number $k$.

\section{The impact of the scattering term}
In this section we mention the impact of the scattering term
in the geodesic equation (Eq.(\ref{eq:eq-a})) due to mass variation
on the CMB angular power spectrum.  This term has been omitted in
the recent literature \cite{Brookfield-b,Zhao:2006zf} (however, it
was correctly included in the earlier work, see
\cite{Anderson:1997un}). Because the term is proportional to
$\frac{\partial m}{\partial x}$ and first order quantity in
perturbation,  our results and those of earlier works
\cite{Brookfield-b,Zhao:2006zf} remain the same in the background 
evolutions. However, as will be shown in the appendix, neglecting
this term violates the energy momentum conservation law at linear
level leading to the anomalously large ISW effect. Because the term
becomes important when neutrinos become massive, the late time ISW
is mainly affected through the interaction between dark energy and
neutrinos. Consequently, the differences show up at large angular
scales. In Fig. (\ref{fig:comparison-cmb-spectrum}), the differences
are shown with and without the scattering term. The early ISW can
also be affected by this term to some extent in some massive
neutrino models and the height of the first acoustic peak could be
changed. However, the position of the peaks stays almost unchanged
because the background expansion histories are the same.
\FIGURE[t]{\epsfig{file=fig1.eps,width=10cm} 
\caption{Differences between the CMB power spectra
with and without the scattering term in the geodesic equation of
neutrinos with the same cosmological parameters.}
\label{fig:comparison-cmb-spectrum} }

\section{Quintessence potentials and Cosmological Constraints}
To determine the evolution of scalar field which couples to neutrinos, we
should specify the potential of the scalar field.
A variety of quintessence effective potentials can be found in the
literature.
In the present paper we examine three type of quintessential potentials.
First we analyze what is a frequently invoked form for the effective
potential of the tracker field, i.e., an inverse power law
originally analyzed by Ratra and Peebles \cite{Ratra:1987rm},
\begin{equation}
V(\phi)=M^{4}\left({M_{pl}\over \phi}\right)^{\alpha}~~~\mbox{(Model I)}~,
\end{equation}
where $V_0=M^4=3 M_{pl}^2 H_0^2 \Omega_{\phi}$ stands for the vacuum
energy, $M_{pl}$ is the plank mass and $\alpha$ is free parameter
which will be constrained from observational data.

We will also consider a modified form of $V(\phi)$ as proposed by
\cite{Brax:1999gp} based on the condition that the quintessence fields
be part of supergravity models. The potential now becomes
\begin{equation}
V(\phi)=M^{4} \left({M_{pl}\over \phi}\right)^{\alpha} e^{3\phi^2/2M_{\rm
pl}^2}~~~\mbox{(Model II)}~,
\end{equation}
where the exponential correction becomes important near the present
time as $\phi \to M_{\rm pl}$. The fact that this potential has a
minimum for $\phi=\sqrt{\alpha/3}M_{\rm pl}$ changes the dynamics.
It causes the present value of $w$ to evolve to a cosmological
constant much quicker than for the bare power-law potential
\cite{Brax:2000yb}.

We will also analyze another class of tracking potential, namely, the
potential of exponential type \cite{Copeland:1997et}:
\begin{equation}
V(\phi) = M^4 e^{-\alpha ({\phi \over M_{pl}})}~~~\mbox{(Model
III)}~,
\end{equation}
This type of potential can lead to accelerating expansion provided that
$\alpha<\sqrt{2}$. In figure (\ref{fig:energy_densities}), we present
examples of evolution of
energy densities with these three types of potentials with vanishing
coupling strength to neutrinos.
\FIGURE[t]{\epsfig{file=fig2.eps,width=10cm} 
\caption{Examples of the evolution of energy density in quintessence and
 the background fields as indicated. Model parameters taken to plot this
 figure are $\alpha=10$, $10$, $1$ for model I, II, III,
 respectively. The other parameters for the dark energy are fixed so that
 the energy densities in three types of dark energy should be the same at
 present.}
\label{fig:energy_densities} }
\subsection{Equation of State in Interacting Dark-Energy with
Neutrinos}
In quintessence models, the scalar field $\phi$ rolls down a
self-interacting potential $V(\phi)$. The equation of state
(EoS) is defined,
\begin{equation}
\omega_{\phi} = { \dot{\phi}^2/2 - V(\phi) \over \dot{\phi}^2/2 +
V(\phi) }. \label{eq:eos}
\end{equation}
It has to satisfy the condition $\omega_{\phi} < -1/3$ for the cosmic
acceleration. Since 
$\omega_{\phi} \geq -1$ from eq.(\ref{eq:eos}), the $\omega_{\phi} <
-1$ regime can not be realized by quintessence. However observations
have shown
exciting possibility of $\omega_{\phi} <-1$, which causes the big-rip
singularity problem by the phantom field at the end of the day. Various 
independent analysis of the {\bf Gold} SN1a data set
\cite{eos-SN1a:2004} indicated it. Present situations of the
observation tell us: First cosmic shear results from the
Canada-France-Hawaii Telescope (CFHT) Wide Synoptic Survey provided
a constraint on a constant equation of state for dark-energy, based
on cosmic shear data alone: $\omega_0 < -0.8 $ with $68 \%$ C.L.
from the Deep Component of the CFHTLS\cite{hoekstra-CFHT:2005}.
 In the Supernova Legacy Survey
(SNLS) \cite{astier-snls:2006} cosmological fits to the first year
SNLS Hubble diagram gave $\omega=-1.023 \pm 0.090 \pm 0.054$ for a
flat cosmology with constant equation of state when combined with
the constraint from the recent Sloan Digital Sky Survey (SDSS)
measurement of baryon acoustic oscillations. WMAP3
data \cite{spergel-wmap3:2006} gave us two different results for
different assumptions: when we assume flat universe including SNLS
data, $\omega=-0.97^{+0.07}_{-0.09}$, however if we drop prior of
flat universe, WMAP + LSS + SNLS data provide
$\omega=-1.062^{+0.128}_{-0.079}$ and
$\Omega_{k}=-0.024^{+0.016}_{-0.013}$. Still observations tell us
the possibility of unexpected $\omega < -1$.

We summarize the possible values of EoS in various quintessence
potential models in table {\ref{table:EoS}}.
\TABLE[t]{
\caption{Equation of state in various quintessence potential models}
\begin{tabular}{@{}c||c||c@{}} \hline
 Quintessence Potentials &
Equation of State (EoS) & References \\ \hline \hline
 $M^{4} \,\, exp(-\lambda
\phi)$ & $\omega=\lambda^2/3-1$  & Ratra \&
Peebles\cite{ratra-peebles:1988}, Wetterich\cite{wetterich:1988} \\
                    & $\lambda =\sqrt{3/\Omega_{\phi}}$, $\Omega_{\phi} < 0.1 -0.15$ &
                    Ferreira \& Joyce\cite{ferreira:1997} \\ \hline
$M^{4+\alpha}/\phi^{\alpha}$, $\alpha > 0$  & $\omega > -0.7$ &
Ratra \& Peebles\cite{ratra-peebles:1988} \\ \hline $M^4 \,
exp(\lambda \phi^2)\, \phi^{-\alpha}$ & $\alpha \geq 11$, $\omega
\simeq -0.82$ & Brax \& Martin\cite{brax-martin:1999-2000} \\
\hline $m^2\phi^2$, \, $\lambda \phi^4$ & PNGB  & Frieman et
al.\cite{frieman:1995} \\
    & $M^4 \, [cos(\phi/f) +1]$  &        \\ \hline
$M^4 \, [exp(M_{pl}/\phi)-1] $ & $\Omega_m >0.2$, $\omega <-0.8$ &
Zlatev, Wang \& Steinhardt\cite{zlatev:1999} \\ \hline $M^4 \,
[cosh(\lambda \phi)-1]^p$ & $p<1/2$, $\omega<-1/3$ & Sahni \&
Wang\cite{sahni-wang:2000} \\ \hline $M^4 \,
sinh^{-\alpha}(\lambda \phi)$ & {\rm early time: inverse power} &
Sahni \& Starobinsky\cite{sahni-starobinsky:2000} \\
    & {\rm late time: exponential} & Urena-L\'{o}pez \&
    Matos\cite{lopez-matos:2000} \\ \hline
$M^4 \, [(\phi-B)^{\alpha} + A] \, exp{(-\lambda \phi)}$ & $\omega
\sim -1$ & Albrecht \& Skordis\cite{albrecht:2000} \\ \hline $M^4
\, exp[\lambda(\phi/M_{pl})^2]$  &  $\omega \sim -1$ & Lee, Olive \&
Pospelov\cite{lee-olive-pospelov:2004} \\
$M^4 \, cosh[\lambda \phi/M_{pl}]$ & $\omega \sim -1$ &  \\ \hline
\hline
\end{tabular} \label{table:EoS}}
In this section we show that $\omega_{\phi} < -1$ naturally arises
when quintessence field interacts with neutrinos. Similar processes
has been done before in the case of interacting dark-energy and
dark-matter \cite{das-khoury:2006}.
\FIGURE[t]{\epsfig{file=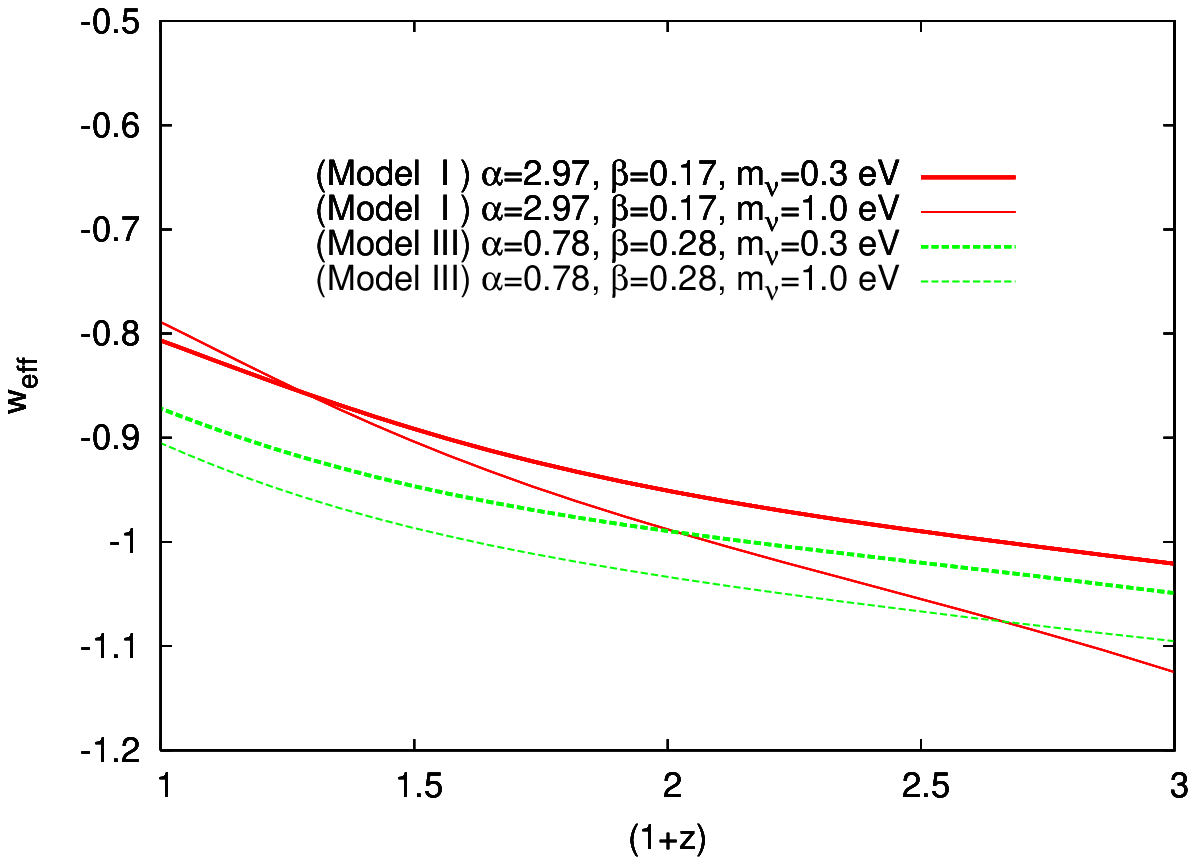,width=10cm}
%
\caption{Time evolution of the effective equation of state parameter
 $w_{\rm eff}$ for models with potentials of inverse power law (Model I)
 and exponential types (Model III). The parameters are fixed to the best
 fitting values except for those shown in the figure. The effective
 equation of state parameter can be smaller than $-1$ at $z>0$.} 
\label{fig:w_eff} }
As pointed out by earlier works it is possible to have the
observational equation of state $w_{\rm eff}$ less than -1 in the
neutrino-dark energy interacting models.  The point is that any
observer would unaware the dark energy interactions, and attribute
any unusual evolution of neutrino energy density to that of dark
energy. This is seen as follows. Let us consider the recent epoch
where the neutrinos have already become massive enough so that the
energy density of neutrinos can be described as
\begin{equation}
\rho_\nu = m_\nu(\phi) n_{\nu}=m_\nu(\phi) n_{\nu,0}/a^3~,
\end{equation}
where $n_\nu,0$ is the number density of neutrinos at present time.
One can decompose this into two parts as
\begin{equation}
\rho_\nu = m_{\nu}(\phi_0)n_{\nu,0}/a^3 +
\left(\frac{m_\nu(\phi)}{m_\nu(\phi_0)}-1\right)m_\nu(\phi_0)n_{\nu,0}/a^3~,
\end{equation}
and hence the Friedmann equation (neglecting baryon and photon
contributions)
\begin{eqnarray}
H^2 &=& \frac{8\pi G}{3}\left(\rho_{\rm
CDM}+\rho_\nu+\rho_\phi\right)
\nonumber \\
&=&\frac{8\pi G}{3}\left(\left(\rho_{\rm
CDM,0}+\rho_{\nu,0}\right)/a^3+\left(\frac{m_\nu(\phi)}{m_\nu(\phi_0)}-1\right)m_\nu(\phi_0)n_{\nu,0}/a^3+\rho_\phi\right)~.
\end{eqnarray}
Therefore, while the first term in the above equation is regarded as
a (total) matter density of our universe, the second and third terms
comprise effective energy density which would be recognized as dark
energy,
\begin{equation}
\rho_{\rm eff} \equiv
\left(\frac{m_\nu(\phi)}{m_\nu(\phi_0)}-1\right)m_\nu(\phi_0)n_{\nu,0}/a^3+\rho_\phi~.
\label{eq:rho_eff}
\end{equation}
In observations one measures the equation of state of dark energy
$w_{\rm eff}$ defined by
\begin{equation}
\frac{d\rho_{\rm eff}}{dt}=-3H(1+w_{\rm eff})\rho_{\rm eff}~.
\end{equation}
Here $w_{\rm eff}$ is related to the equation of state of
quintessence $w_\phi$ through
\begin{eqnarray}
w_{\rm eff}&=&\frac{w_\phi}{1-x}~,\\
x &=&
\left(\frac{m_\nu(\phi)}{m_\nu(\phi_0)}-1\right)\frac{m_\nu(\phi_0)n_{\nu,0}/a^3}{\rho_\phi}~,
\end{eqnarray}
which are derived from Eqs.(\ref{eq:Qddot}), (\ref{eq:rho_phi}), and
(\ref{eq:rho_eff}). An example of time evolution of $w_{\rm eff}$ is
depicted in Fig. (\ref{fig:w_eff}). 

\subsection{Time evolution of neutrino mass and energy density in
scalar
 field}
For an illustration we also plot examples of evolution of energy
densities for interacting case with inverse power law potential (Model
I) in Fig. (\ref{fig:energy_densities_coupled}).
In interacting dark energy cases, the evolution of the scalar field is
determined both by its own potential and interacting term from neutrinos.
When neutrinos are highly relativistic, the interaction term can be expressed
as
\begin{equation}
(\rho_\nu-3P_\nu)\approx
 \frac{10}{7\pi^2} (am_\nu)^2 \rho_{\nu_{\rm massless}}~,
\end{equation}
where $\rho_{\nu_{\rm massless}}$ denotes the energy density of
neutrinos with no mass. The term roughly scales as $\propto a^{-2}$,
and therefore, it dominates deep in the radiation dominated era.
However, because the motion of the scalar field driven by this
interaction term is almost suppressed by the friction term, $-2{\cal
H}\dot\phi$. The scalar field satisfies the slow roll condition
similar to the inflation models, $-2{\cal H}\dot\phi\approx
a^2\frac{\partial m_\nu}{\partial \phi}(\rho_\nu-3P_\nu)$. Thus, the
energy density in scalar field and the mass of neutrinos is frozen
there. 
These behaviors are clearly seen in Figs.
(\ref{fig:energy_densities_coupled}) and (\ref{fig:mass_evolution}).
The derivation of the analytic expression for the behavior of
varying neutrino mass in early time of the universe is given at
Appendix B.
\FIGURE[t]{\epsfig{file=fig4.eps,width=10cm}
\caption{Examples of the evolution of energy density in quintessence and
 the background fields in coupled cases with inverse power law potential
 (Model I). Model parameters taken to plot this
 figure are $\alpha=1$, $\beta=1$, $3$ as indicated. The other
 parameters for the dark energy are fixed so that
 the energy densities in three types of dark energy should be the same at
 present.}
\label{fig:energy_densities_coupled} }
\FIGURE[t]{\epsfig{file=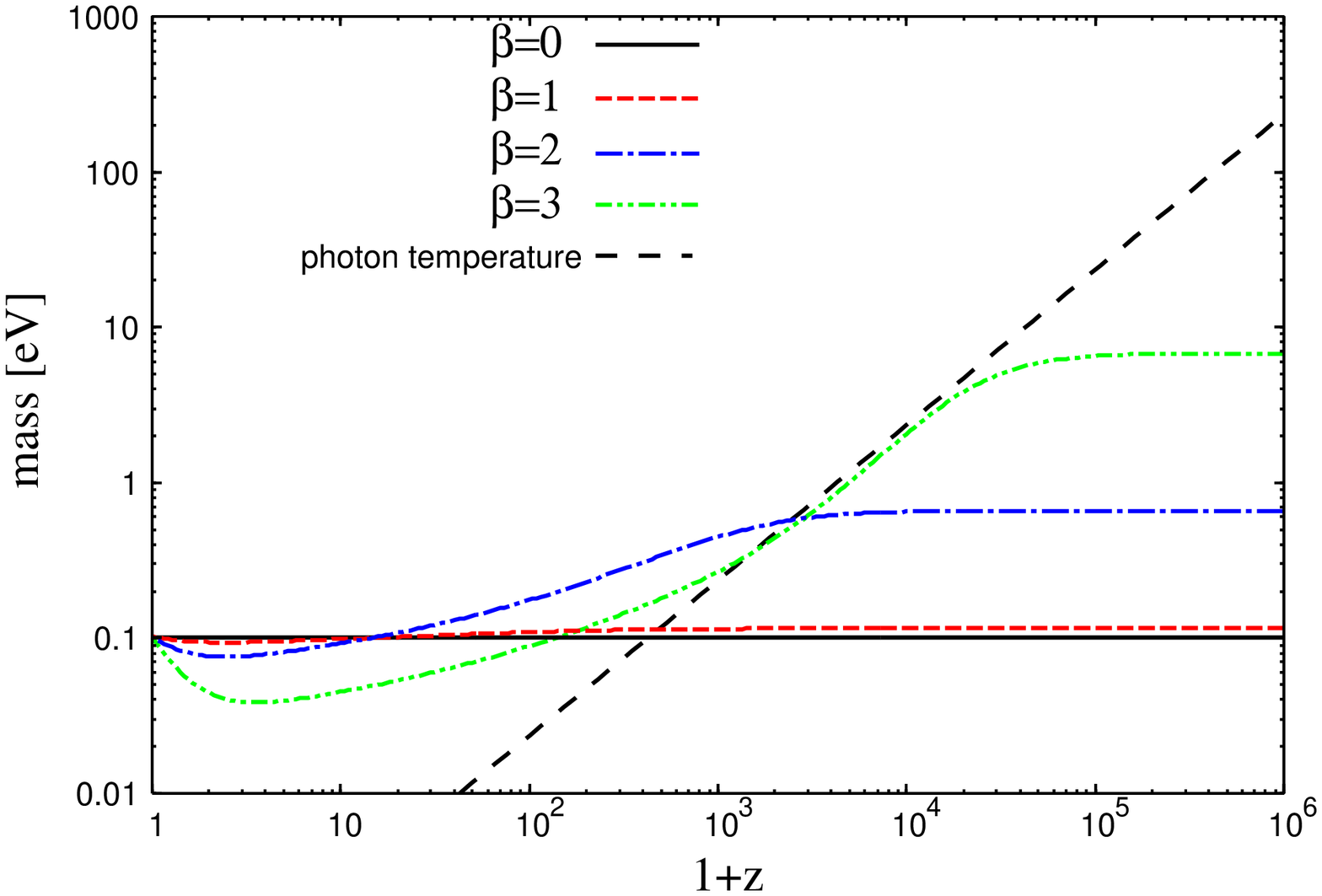,width=10cm}
\caption{Examples of the time evolution of neutrino mass in power law
 potential models (Model I) with $\alpha=1$ and $\beta=0$ (black solid line),
 $\beta=1$ (red dashed line), $\beta=2$ (blue dash-dotted line),
 $\beta=3$ (dash-dot-dotted line). The larger coupling parameter leads
 to the larger mass in the early universe.
}
\label{fig:mass_evolution} }
\subsection{Constrains on the MVN parameters and Neutrino Mass Bound}

\subsubsection{Neutrino Mass Bounds from Beta
Decays and Large Scale Structures}

In this section, we review in brief the present status of the
determination of absolute neutrino mass from beta decay experiments
and cosmological observation data. The existence of the tiny
neutrino masses qualifies as the first evidence of new physics
beyond the Standard Model. The answer to the hot questions on (1)
whether neutrinos are Dirac or Majorana fermions ? (2) what kind of
mass hierarchy pattern they have ? (3) what are the absolute values
of the neutrino masses, will provide us the additional knowledge
about the precise nature of this new physics, and in turn about the
nature of new forces beyond the Standard Model. There are three well
known ways to get the direct information on the absolute mass of
neutrinos by using: Tritium $\beta$-decay experiment, neutrinoless
double beta decay experiment, and astrophysical observations.

The standard method for the measurement of the absolute value of the
neutrino mass is based on the detailed investigation of the
high-energy part of the $\beta$-spectrum of the decay of tritium:
\begin{equation}
{}^3H \longrightarrow {}^3He + e^{-} + \bar{\nu}_e
\end{equation}
This decay has a small energy release ($E_0 \simeq 18.6 keV$) and a
convenient life time ($T_{1/2} = 12.3 $ years). Since the flavor
eigenstates are different from mass eigenstates in neutrino sector,
in general, electron neutrino can be expressed as
\begin{equation}
\nu_{eL} = \sum_{i} U_{ei} \, \nu_{iL},
\end{equation}
where $\nu_i$ is the field of neutrino with mass $m_i$, and U is the
unitary mixing matrix. Neglecting the recoil of the final nucleus,
the spectrum of the electrons is given:
\begin{equation}
{d\Gamma \over d E} = \sum_{i} |U_{ei}|^2 \,\, {d\Gamma_{i} \over
dE},
\end{equation}
and the resulting spectrum can be analyzed in term of a single
mean-squared electron neutrino mass
\begin{equation}
\langle m_{\beta} \rangle^2 =\sum_{j} m_{j}^2|U_{ej}|^2 =
m_1^2|U_{e1}|^2 + m_2^2|U_{e2}|^2 + m_3^2 |U_{e3}|^2
\end{equation}
If the neutrino mass spectrum is practically degenerate: $m_1 \simeq
m_2 \simeq m_3$, the neutrino mass can be measured in these
experiments. Present-day tritium experiments Mainz\cite{mainz} and
Troitsk\cite{troitsk} gave the following results:
\begin{eqnarray}
m_1^2 &=& (-1.2\pm 2.2 \pm 2.1) \, eV^2 \hspace{5mm} {\rm (Mainz)},
\\
&=& ( -2.3 \pm 2.5 \pm 2.0) \, eV^2 \hspace{5mm}  {\rm (Troitsk)}.
\end{eqnarray}
This value corresponds to the upper bound
\begin{equation}
m_1 < \, 2.2 eV  \hspace{3mm} (95 \% C.L.)
\end{equation}

Another useful method is by using the neutrinoless double beta
decay. The search for neutrinoless double $\beta$-decay
\begin{equation}
(A,Z) \longrightarrow (A,Z_2) + e^- + e^-
\end{equation}
for some even-even nuclei is the most sensitive and direct way of
investigating the nature of neutrinos with definite masses. In this
process, total lepton number is violated and is allowed only if the
massive neutrinos are Majorana particles. The rate of
$0\nu\beta\beta$ is approximately
\begin{equation}
{1 \over T^{0\nu}_{1/2}} = G_{0\nu}(Q_{\beta\beta}, Z) \,
|M_{0\nu}|^2 \,\, \langle m_{\beta\beta} \rangle^2,
\end{equation}
where $G^{0\nu}$ is the phase space factor for the emission of the
two electrons, $M_{0\nu}$ is nuclear matrix elements, and
$<m_{\beta\beta}> $ is the effective Majorana mass of the electron
neutrino:
\begin{equation}
\langle m_{\beta\beta} \rangle \equiv |\sum_{i} U_{ei}^2 m_i|
\label{eq:mbetabeta}
\end{equation}
We can write eq.(\ref{eq:mbetabeta}), for normal and inverted
hierarchy respectively, in terms of mixing angles and
$\Delta_s^2=m_2^2 -m_1^2 = 7.9 ^{+2.8}_{-2.9} \cdot 10^{-5} \, eV^2$, 
$\Delta_a= \pm (m_3^2 -m_2^2) = \pm(2.6 \pm 0.2) \cdot 10^{-3} \, eV^2$ 
at the $3\sigma$ level and CP
phases as follows:
\begin{eqnarray}
\langle m_{ee} \rangle &=& \left|c_2^2 c_3^2 m_1 + c_2^2 s_3^2
e^{i\phi_2} \sqrt{\Delta_s^2 + m_1^2} + s_2^2 e^{i\phi_3}
\sqrt{\Delta_a^2 +
m_1^2} \right|, \hspace{1mm} {\rm (normal \,\, hierarchy)} \\
\langle m_{ee} \rangle &=& \left|s_2^2 m_1 + c_2^2 s_3^2 e^{i\phi_2}
\sqrt{\Delta_a^2 - \Delta_s^2 + m_1^2} + c_2^2s_2^2 e^{i\phi_3}
\right|, \hspace{6mm} {\rm (inverted \,\, hierarchy)}.
\end{eqnarray}
However, $0\nu\beta\beta$ decay have not yet been seen experimentally. The
most stringent lower bounds for the life-time of
$0\nu\beta\beta$-decay were obtained in the
Heidelberg-Moscow\cite{heidelberg-moscow} and IGEX\cite{igex}
${}^{76} Ge$ experiments:
\begin{eqnarray}
&& T^{0\nu}_{1/2} \geq 1.9 \cdot 10^{25} years \hspace{5mm} (90 \%
C.L.) \hspace{5mm} {\rm Heidelberg-Moscow}, \\
&& T^{0\nu}_{1/2} \geq 1.57 \cdot 10^{25} years \hspace{5mm} (90 \%
C.L.) \hspace{5mm} {\rm IGEX}.
\end{eqnarray}
Taking into account different calculation of the nuclear matrix
elements, from these results the following upper bounds were
obtained for the effective Majorana mass:
\begin{equation}
|m_{\beta\beta}| \, < \, (0.35 - 1.24) \, eV
\end{equation}
Many new experiments (including CAMEO,CUORE,COBRA, EXO, GENIUS,
MAJORANA, MOON and XMASS experiments) on the search for the
neutrinoless double $\beta$-decay are in preparation at present. In
these experiments the sensitivities
\begin{equation}
|m_{\beta\beta}| \simeq (0.1 - 0.015)\, eV
\end{equation}
are expected to be achieved. More detail discussions will be appeared in
the separated paper\cite{yyk-OMEG07}.

Within the standard cosmological model, the relic abundance of
neutrinos at present epoch was come out straightforwardly from the
fact that they follow the Fermi-Dirac distribution after freeze
out, and their temperature is related to the CMB radiation
temperature $T_{CMB}$ today by $T_{\nu} =(4/11)^{1/3} T_{CMB}$ with
$T_{CMB}=2.726$ K, providing
\begin{equation}
n_{\nu} = { 6 \zeta(3) \over 11 \pi^2} \, T_{CMB}^3,
\label{eq:nu-density}
\end{equation}
where $\zeta(3)\simeq 1.202$, which gives $n_{\nu} \simeq 112
cm^{-3}$ for each family of neutrinos at present. By now the massive
neutrinos become non-relativistic, and their contribution to the
mass density ($\Omega_{\nu}$) of the universe can be expressed as
\begin{equation}
\Omega_{\nu} h^2 ={\Sigma \over 93.14 eV}. \label{eq:omega-nu}
\end{equation}
where $\Sigma$ stands for the sum of the neutrino masses.
In this relation, the effect of three neutrino 
oscillation is included \cite{Magano:2005}.
We should notice that when obtaining the limit of neutrino masses
one usually assumes:
\begin{itemize}
\item the standard spatially flat $\Lambda CDM$ model with adiabatic
primordial perturbations,
\item they have no non-standard interactions,
\item neutrinos decoupled from the thermal background at the
temperatures of order 1 MeV.
\end{itemize}
These simple conditions can be modified from several effects: due
to a sizable neutrino-antineutrino asymmetry, due to additional
light scalar field coupled with neutrinos \cite{beacom:2004}, and due
to the light sterile neutrino \cite{dodelson:2006}. However, analysis
of WMAP and 2dFGRS data gave independent evidence for small lepton
asymmetries \cite{{hannestad:2003},{pierpaoli:2003}}, and  such a
scenario with a light scalar field  is strongly disfavored by the
current CMB power spectrum data \cite{hannestad:2005}. We will not
therefore take into account such non-standard couplings of neutrinos
in the following. In addition, current cosmological observations are
sensitive to neutrino masses $0.1 \,{\rm eV} \, < \, \Sigma \, <\,
2.0 \, {\rm eV}$. In this mass scale, the mass-square differences
are small enough and all three active neutrinos are nearly
degenerate in mass. Therefore we take the assumption of degenerate
mass hierarchy. Even if we consider different mass hierarchy
pattern, it will be very difficult to distinguish such hierarchy
patterns from cosmological data alone
\cite{slosar:2006}.

After neutrinos decoupled from the thermal background, they stream freely
and their density perturbations are damped on scale smaller than their
free streaming scale. Consequently the perturbations of cold 
dark matter (CDM) and baryons grow more slowly because of the missing
gravitational contribution from neutrinos. The free streaming scale
of relativistic neutrinos grows with the Hubble horizon. When the
neutrinos become non-relativistic, their free streaming scale shrinks,
and they fall back into the potential wells. The neutrino density
perturbation with scales larger than the free streaming scale resumes to
trace those of the other species.  
Thus the free streaming effect suppresses the power spectrum on scales
smaller than the horizon when the neutrinos become
non-relativistic. 
The co-moving wave number corresponding to this scale is given by
\begin{equation}
k_{nr} = 0.026 \,\left( {m_{\nu} \over 1 \, eV} \right)^{1/2} \,
\Omega_m^{1/2} \, h \, {\rm Mpc}^{-1}, \label{jeans-length}
\end{equation}
for degenerated neutrinos, with almost same mass $m_{\nu}$. The
growth of Fourier modes with $k > k_{nr}$ will be suppressed because
of neutrino free-streaming. The power spectrum of matter
fluctuations can be written as
\begin{equation}
P_m(k,z) = P_{*}(k) \, T^2(k,z), \label{eq:power-spectrum}
\end{equation}
where $P_{*}(k)$ is the primordial spectrum of matter fluctuations,
to be a simple power law $P_{*}(k) = A \, k^n$, where A is the
amplitude and n is the spectral index. 
Here the transfer function $T(k,z)$ represents the evolution of
perturbation relative to the largest scale. If some fraction of the
matter density (e.g., neutrinos or dark energy) is unable to cluster,
the speed of growth of perturbation 
becomes slower. 
Because the contribution to the fraction of matter density from
neutrinos is proportional to their masses (Eq. (\ref{eq:omega-nu})), 
the larger mass leads to the smaller growth of perturbation.
The suppression of the power spectrum on small scales is
roughly proportional to $f_{\nu}$ \cite{Hu:1997mj}:
\begin{equation}
{ \Delta P_m(k) \over P_m(k)} \simeq -8 f_{\nu}.
\label{eq:power-spectrum-02}
\end{equation}
where $f_{\nu}=\Omega_{\nu}/\Omega_{M}$ is the fractional
contribution of neutrinos to the total matter density. This result
can be understood qualitatively from the fact that only a fraction
$(1-f_{\nu})$ of the matter can cluster when massive neutrinos are
present \cite{silk:1980}.
\TABLE[t]{
\caption{Recent cosmological neutrino mass bounds ($95 \%$ C.L.)}
\begin{tabular}{@{}c||c||c@{}} \hline
Cosmological Data Set & $\Sigma$ bound ($2 \sigma$) & References \\
\hline \hline
CMB (WMAP-3 year alone) & $< \, 2.0$ eV & Fukugita et al.\cite{fukujita:2006} \\
LSS[2dFGRS] & $ < \, 1.8 $ eV & Elgaroy et al.\cite{elgaroy:2002} \\
CMB + LSS[2dFGRS] & $< \, 1.2$ eV & Sanchez et al.\cite{sanchez:2005} \\
              "        &  $< \, 1.0 $ eV  &
              Hannestad\cite{hannestad:2004} \\
CMB + LSS + SN1a & $< \, 0.75 $ eV & Barger et al.\cite{barger:2003} \\
       "            & $< \, 0.68 $ eV & Spergel et al.\cite{spergel:2006} \\
CMB + LSS + SN1a + BAO & $< \, 0.62 $ eV & Goobar et al.\cite{goobar:2006} \\
        "               & $< \, 0.58 $ eV &                              \\
CMB + LSS + SN1a + Ly-$\alpha$ & $< \, 0.21 $ eV &    Seljak et al.\cite{seljak:0604335} \\
CMB + LSS + SN1a + BAO + Ly-$\alpha$ & $ < \, 0.17 $ eV & Seljak et
al.\cite{seljak:0604335} \\ \hline
 \hline
\end{tabular} \label{table:neutrino-mass-bound}}

Analyses of CMB data are not sensitive to neutrino masses 
if neutrinos behave as massless particles at the epoch of last scattering.
According to the analytic
consideration in \cite{ichikawa:2005}, since the redshift when
neutrino becomes non-relativistic is given by $1+z_{nr}=6.24 \cdot
10^4 \, \Omega_{\nu} \, h^2$ and $z_{rec}=1088$, neutrinos become
non-relativistic before the last scattering 
when $\Omega_{\nu}h^2 > 0.017$ (i.e. $\Sigma > 1.6 e V$). 
Therefore the dependence of the position of the first peak
and the height of the first peak on $\Omega_{\nu}h^2$ has a
turning point at $\Omega_{\nu}h^2 \simeq 0.017$. This value also
affects CMB anisotropy via the modification of the integrated
Sachs-Wolfe effect due to the massive neutrinos. 
However an important role of CMB data
is to constrain  other parameters that are degenerate with $\Sigma$.
Also, since there is a range of scales common to the CMB and LSS
experiments, CMB data provides an important constraint on the bias
parameters. We summarize some of the recent cosmological neutrino
mass bounds in table \ref{table:neutrino-mass-bound}.

\subsubsection{Neutrino Mass Bound in Neutrino-Dark Energy Model}
\begin{center}
\begin{figure}[t]
\vspace{0cm} \epsfxsize=5cm \centerline{
\rotatebox{0}{\includegraphics[width=0.7\textwidth]{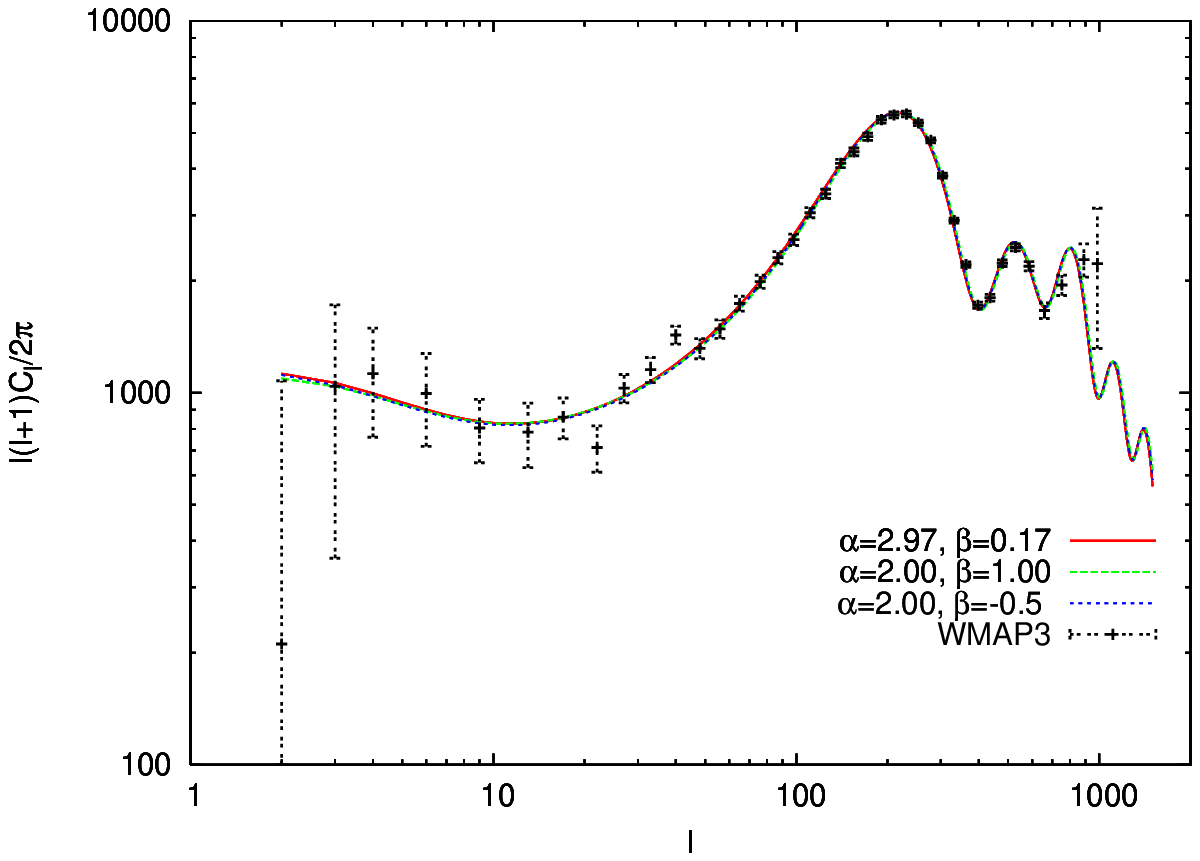}} }
\vspace{0cm} \epsfxsize=5cm \centerline{
\rotatebox{0}{\includegraphics[width=0.7\textwidth]{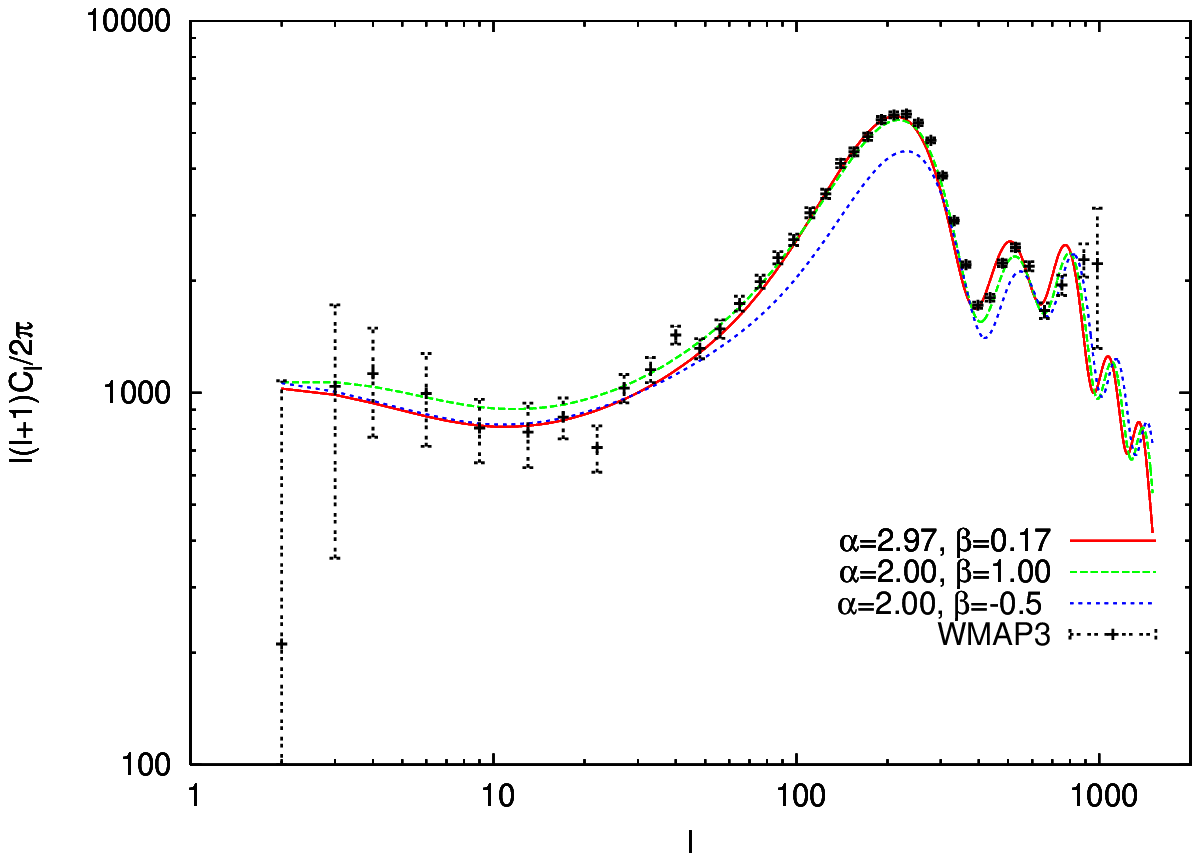}} }
\epsfxsize=5cm 
\caption{The CMB angular power spectra for Model I with different neutrino masses:
 $M_\nu=0.3$ eV (upper panel) and $M_\nu=1.0$ eV (down panel).
The solid line is
the best fit for the model ($(\alpha,\beta)=(2.97,0.170)$), 
the other lines are models with different
 parameter value of $\alpha$ and $\beta$ as indicated.
The points are WMAP three year data.
 \label{fig:IPL_cl_spectra} }
\end{figure}
\end{center}

As was shown in the previous sections, the coupling between
cosmological neutrinos and dark energy quintessence could modify the
CMB and matter power spectra significantly.  It is therefore
possible and also important to put constraints on coupling
parameters from current observations. For this purpose, we use the
WMAP3 \cite{Hinshaw:2006ia,Page:2006hz} and 2dFGRS \cite{Cole:2005sx}
data sets.
In figure \ref{fig:IPL_cl_spectra}, we demonstrate, as an example, the total neutrino mass
contribution on the CMB angular power spectra with the inverse power law potential (Model 1).
We can see that the deviations from observation data becomes severe 
when the total neutrino mass increases
The flux power spectrum of the Lyman-$\alpha$ forest can be used to
measure the matter power spectrum at small scales around $z\la 3$ \cite{McDonald:1999dt,Croft:2000hs}.
It has been shown, however, that the resultant constraint on neutrino
mass can vary significantly from $\sum m_\nu < 0.2$eV to $0.4$eV
depending on the specific Lyman-$\alpha$ analysis used \cite{Goobar:2006xz}.
The complication arises because the result suffers from the
systematic uncertainty regarding to the model for the intergalactic
physical effects, i.e., damping wings, ionizing radiation fluctuations,
galactic winds, and so on \cite{McDonald:2004xp}.
Therefore, we conservatively omit the Lyman-$\alpha$ forest data from
our analysis.

Because there are many other cosmological parameters than the MVN
parameters, we follow the Markov Chain Monte Carlo(MCMC) global fit
approach \cite{MCMC} to explore the likelihood space and marginalize
over the nuisance parameters to obtain the constraint on
parameter(s) we are interested in 
\cite{WHEPP8:2004}.
Our parameter space consists of
\begin{equation}
\vec{P}\equiv (\Omega_bh^2,\Omega_ch^2,H,\tau,A_s,n_s,m_i,\alpha,\beta)~,
\end{equation}
where $\Omega_bh^2$ and $\Omega_ch^2$ are the baryon and CDM densities
in units of critical density, $H$ is the Hubble parameter, $\tau$ is the
optical depth of Compton scattering to the last scattering surface, $A_s$
and $n_s$ are the amplitude and spectral index of primordial density
fluctuations, and $(m_i,\alpha,\beta)$ are the parameters of MVN
defined in section III. We have put priors on MVN parameters as
$\alpha>0$, and $\beta>0$ for simplicity and saving the computational time.
\FIGURE[t]{\epsfig{file=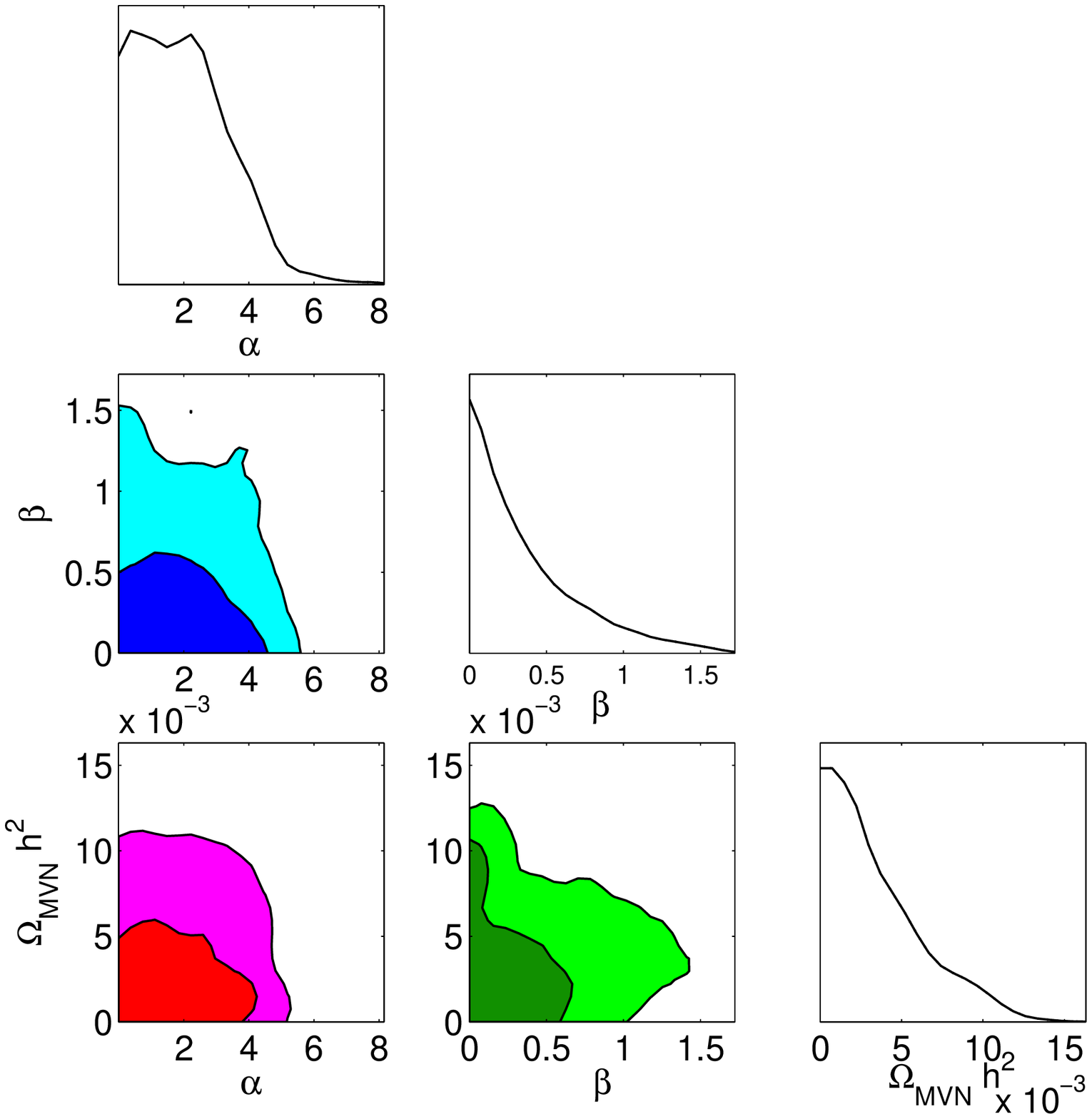,width=10cm}
\caption{Contours of constant relative probabilities in two dimensional
 parameter planes for inverse power law models. Lines correspond to 68\% and 95.4\% confidence limits.}
\label{fig:ratra_tri} }
\FIGURE[h]{\epsfig{file=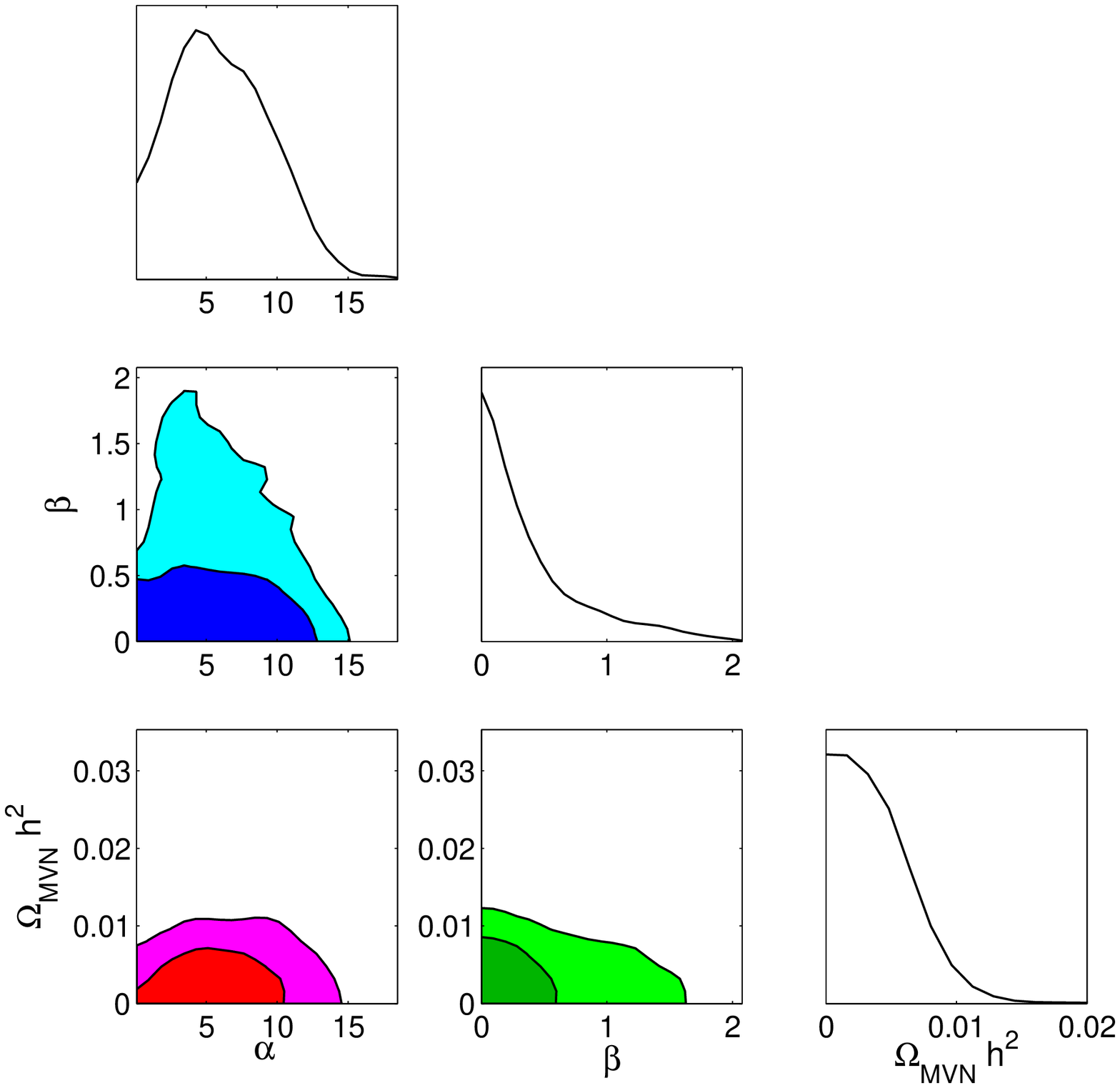,width=10cm}
\caption{Same as Fig.(\ref{fig:ratra_tri}), but for SUGRA type models.
}
\label{fig:SUGRA_tri} }
\FIGURE[h]{\epsfig{file=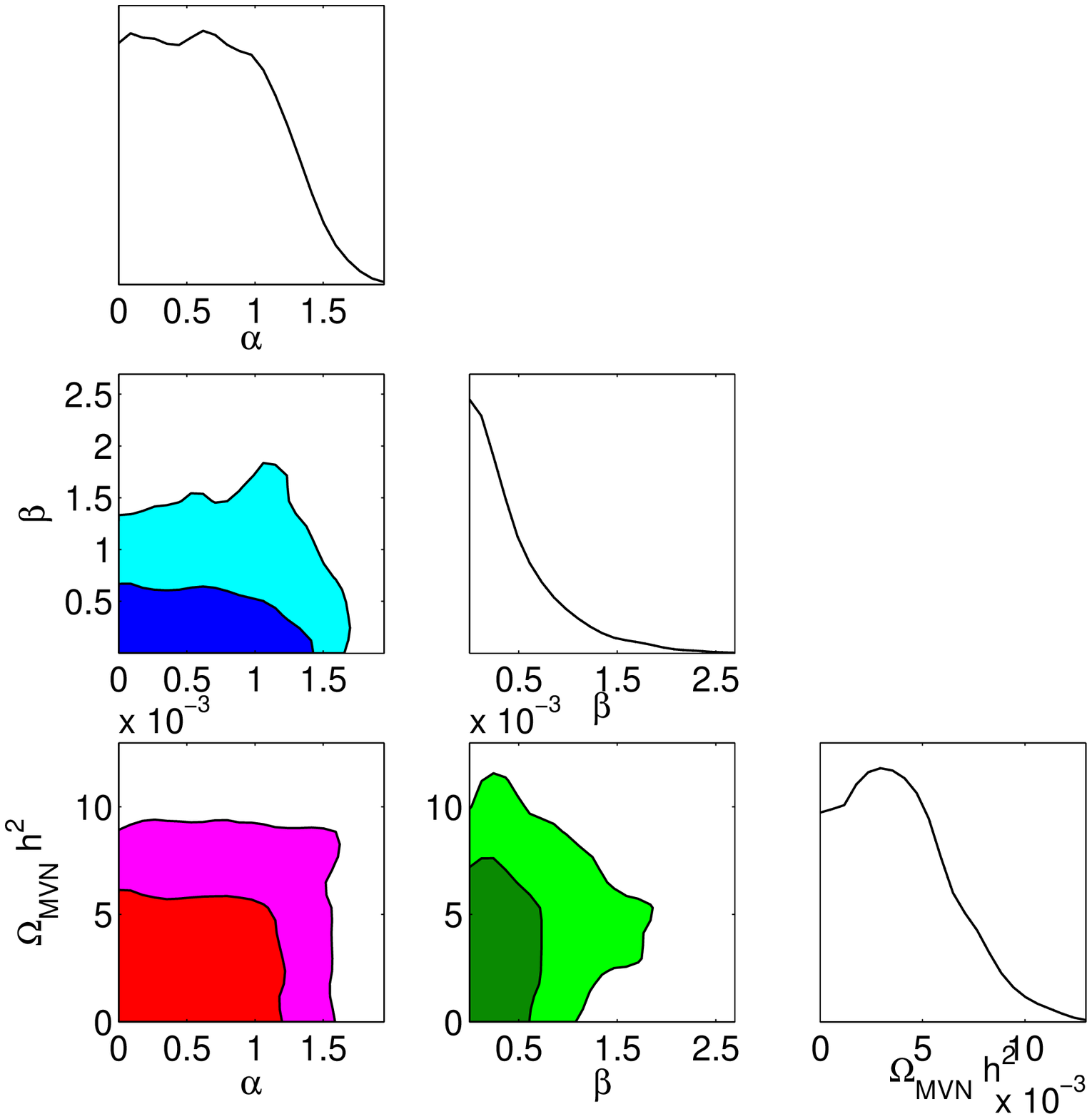,width=10cm}
\caption{Same as Fig.(\ref{fig:ratra_tri}), but for exponential type models.
}
\label{fig:EXP_tri} }

Our results are shown in Figs.(\ref{fig:ratra_tri}) -
(\ref{fig:EXP_tri}).
In these figures we do not observe the strong degeneracy between the
introduced parameters. This is why one can put tight constraints on
MVN parameters from observations. For both models we consider, larger
$\alpha$ leads larger $w$ at present. Therefore large
$\alpha$ is not allowed due to the same reason that larger $w$ is not
allowed from the current observations.

On the other hand, larger $\beta$ will generally lead larger $m_\nu$ in
the early universe. This means that the effect of neutrinos on the
density fluctuation of matter becomes larger leading to the larger
damping of the power at small scales. A complication arise because the
mass of neutrinos at the transition from the ultra-relativistic regime
to the non-relativistic one is not a monotonic function of $\beta$ as
shown in Fig.(\ref{fig:mass_evolution}). Even so, the coupled neutrinos give larger decrement
of small scale power, and therefore one can limit the coupling parameter
from the large scale structure data.

One may wonder why we can get such a tight constraint on $\beta$,
because it is naively expected that large $\beta$ value should be
allowed if $\Omega_\nu h^2 \sim 0$. In fact, a goodness of fit is still
satisfactory with large $\beta$ value when $\Omega_\nu h^2 \sim
0$. However, the parameters which give us the best goodness of fit
does not mean the most likely parameters in general. In our parametrization, the
accepted total volume by MCMC in the parameter space where $\Omega_\nu h^2
\sim 0$ and $\beta\ga 1$ was small, meaning that the probability of such a
parameter set is low.
\begin{center}
\begin{figure}
\vspace{0cm} \epsfxsize=5cm \centerline{
\rotatebox{0}{\includegraphics[width=0.7\textwidth]{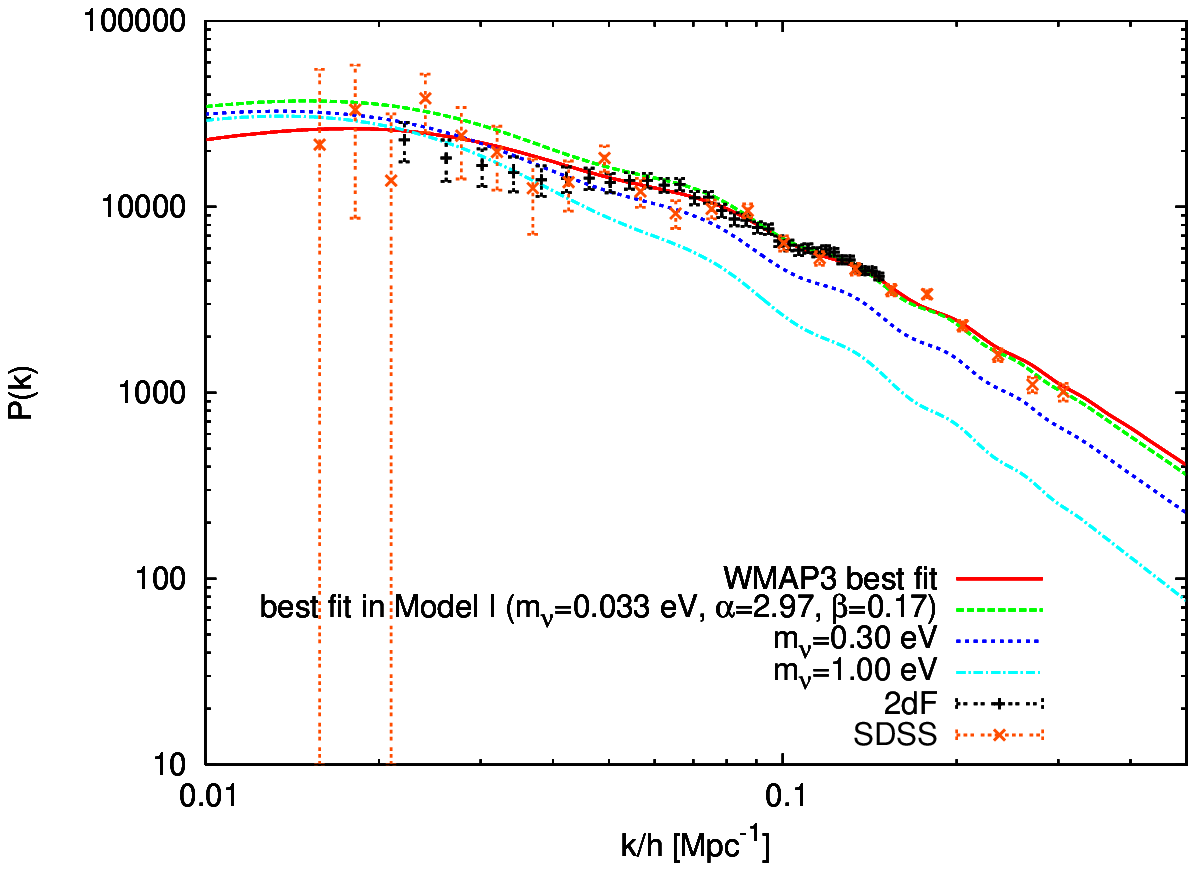}} }
\vspace{0cm} \epsfxsize=5cm \centerline{
\rotatebox{0}{\includegraphics[width=0.7\textwidth]{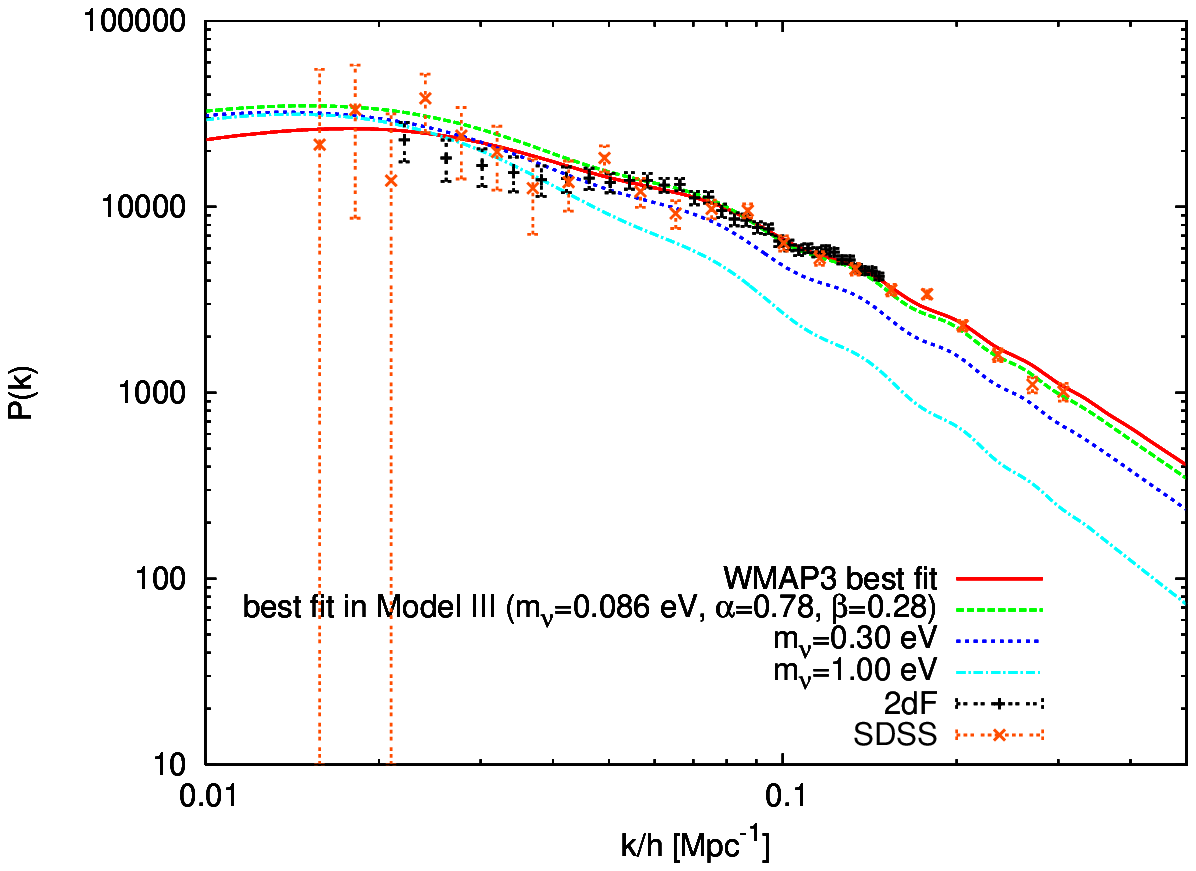}} }
\epsfxsize=5cm \caption{Examples of the total mass contributions in
the matter power spectrum in Model I (upper panel) and Model III
(down panel). For both panels we plot the best fitting lines (green
dashed), lines with larger neutrino masses $M_\nu=0.3$ eV (blue
dotted) and $M_\nu=1.0$ eV (cyan dot-dashed) with the other
parameters fixed to the best fitting values. Note that while lines
with $M_\nu=0.3$ eV can fit to the data well by arranging the other
cosmological parameters, lines with $M_\nu=1.0$ eV can not.
 \label{fig:nu-mass-PS} }
\end{figure}
\end{center}
In tables \ref{tab1}-\ref{tab3}, we summerize our results of global analysis within
$1\sigma$ and $2\sigma$ deviations for different types of quintessence potential 
with WMAP3 years data.
We find no observational signature which favors the coupling between
Mass Varying Neutrinos and quintessence scalar field, and obtain the upper limit on the
coupling parameter as
\begin{equation}
\beta < 0.46, 0.47, 0.58 \,\,(1 \,\sigma); \,\, 
[1.11,~ 1.36,~ 1.53 \,\,(2 \,\sigma)],
\end{equation}
and the present mass of neutrinos is also limited to
\begin{equation}
\Omega_\nu h^2_{\rm{today}} < 0.0044, ~0.0048, ~0.0048~ \,\,(1\,\sigma); \,\,
 [0.0095,~ 0.0090,~ 0.0084~ \,\, (2\,\sigma)],
\end{equation}
for models I, II and III, respectively.
When we apply the relation
between the total sum of the neutrino masses $M_{\nu}$ and their
contributions to the energy density of the universe:
$\Omega_{\nu}h^2=M_{\nu}/(93.14 eV)$, we obtain the constraint on
the total neutrino mass: $M_{\nu} < 0.45~ eV (68 \% C.L.) \,\,[0.87~ eV (95 \% C.L.)$] in the
neutrino probe dark-energy model. The total neutrino mass
contributions in the power spectrum is shown in Fig
\ref{fig:nu-mass-PS}, where we can see the significant deviation
from observation data in the case of  large neutrino masses.

\begin{figure}
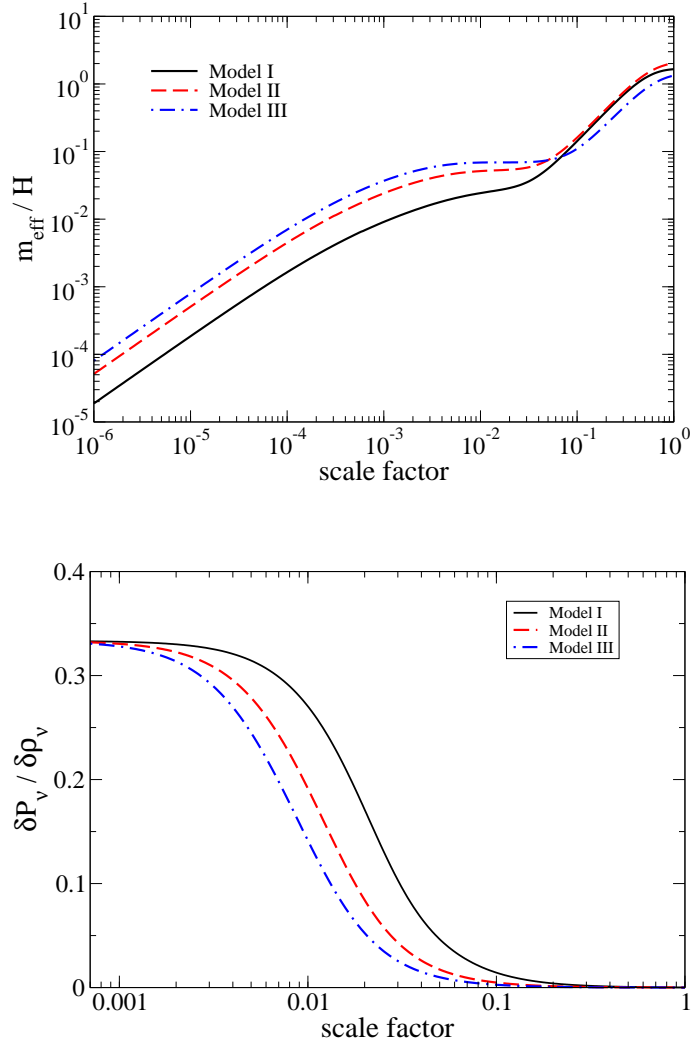

\vspace{0cm} \epsfxsize=5cm \centerline{
\rotatebox{0}{\includegraphics[width=0.6\textwidth]{meffoH.eps}} }
\vspace{1cm} \epsfxsize=5cm \centerline{
\rotatebox{0}{\includegraphics[width=0.6\textwidth]{cs2_nu.eps}} }
\caption{(upper panel): Typical evolution of the effective mass of the
 quintessence scalar field relative to the Hubble scale, for all models
 considered in this paper. (down panel): Typical evolution of the sound speed
 of neutrinos $c_s^2 = \delta P_\nu/\delta\rho_\nu$ with the wave number
 $k=2.3\times 10^{-3}$ Mpc$^{-1}$, for models as
 indicated. The values stay 
 positive stating from $1/3$ (relativistic) and neutrinos are stable
 against the density fluctuation. 
\label{fig:m_eff} }
\end{figure}

Before concluding the paper we should comment on the stability issue
in the present models. As shown in \cite{Afshordi:2005ym,Bean:2007ny}, 
some class of models with mass
varying neutrinos suffers from the adiabatic instability at the
first order perturbation level. This is caused by an additional force
on neutrinos mediated by the quintessence scalar field and occurs when
its effective mass is much larger than the Hubble horizon scale, where
the effective mass is defined by $m_{\rm eff}^2={d^2 V_{\rm
eff}}/{d\phi^2}$.  To
remedy this situation one should consider an appropriate quintessential
potential which has a mass comparable the horizon scale at present, and
the models considered in this paper are the case
\cite{Brookfield-b}. Interestingly, some authors have found that one can
construct viable MVN models by choosing certain couplings and/or
quintessential potentials
\cite{Bjaelde:2007ki,Kaplinghat:2006jk,Takahashi:2006jt}. Some 
of these models even realises $m_{\rm eff} \gg H$. In  
Fig.(\ref{fig:m_eff}), masses of the scalar  
field relative to the horizon scale $m_{\rm eff}/H$ are plotted. We find that $m_{\rm eff}<H$ for almost all period and
the models are stable. We also depict in Fig.(\ref{fig:m_eff}) the sound
speed of neutrinos defined by $c^2_s = \delta P_\nu/\delta\rho_\nu$ with
a wave number $k=2.3\times 10^{-3}$ Mpc$^{-1}$. 

\TABLE[t]{
\caption{Global analysis data within $1\sigma$ and $2\sigma$
deviations for the inverse power law type (Model-I) of the
quintessence potential. (*)Using WMAP data alone, they found
1.93, but along with either SDSS or 2dFGRS galaxy redshift data,
they have 1.83 or 0.97 at $95\%$ confidence.} 
\begin{tabular}{@{}c||c|c||c|c||c@{}}
\hline Quantites &
Mean & SDDEV &  $1\sigma$-ranges & $2\sigma$-ranges & WMAP3 data ($\Lambda$CDM)  \\
\hline
$\alpha$ & $2.08$ & $1.35$ & $< 2.63$ & $< 4.38$ & ---       \\
$\beta$  & $0.38$ &$0.35$ & $< 0.46 $ & $< 1.12$ & ---       \\
$\Omega_B\, h^2[10^2]$ & $2.21$ & $0.07$ & $2.15 - 2.28$ & $2.09 - 2.36$ & $2.23\pm 0.07$  \\
$\Omega_{CDM}\, h^2[10^2]$ & $11.09$ & $0.62$ & $10.52-11.68$ & $9.87-12.30$ & $12.8\pm 0.8$ \\
$H_0$ & $65.97$ & $3.61$ & $62.30 - 69.37$ & $58.39 - 72.10$ & $72\pm 8$  \\
$Z_{re}$ & $10.87$ & $2.58$ & $9.81 - 12.15$ & $6.13 - 14.94$ &   ---     \\
$n_s$ & $0.95$ & $0.02$ & $0.94 - 0.97$ & $0.92 - 0.99$ & $0.958\pm 0.016$      \\
$A_s[10^{10}]$ & $20.66$ & $1.31$ & $19.38 - 21.92$ & $18.25 - 23.41$ & ---- \\
$\Omega_{Q}[10^2]$ & $68.54$ & $4.81$ & $64.02-72.94$ & $57.43-75.60$ & $71.6\pm 5.5$      \\
$Age/Gyrs$ & $13.95$ & $0.20$ & $13.76-14.15$ & $13.59-14.40$ &  $13.73\pm 0.16$ \\
$\Omega_{\nu}\,h^2[10^2]$ &$0.36$ & $0.29$ & $< 0.44$ & $< 0.95$ &
$< 1.93 (95\% C.L.)^{*}$ \\
$\tau$ & $0.084$ & $0.029$ & $0.055- 0.112$ & $0.031-0.143$ & $0.089
\pm 0.030$  \\ \hline
\end{tabular} \label{tab1}}
\TABLE[t]{
\caption{Global analysis data within $1\sigma$ and $2\sigma$
deviations for the SUGRA type (Model-II) of the quintessence
potential.} 
\begin{tabular}{@{}c||c|c||c|c||c@{}} \hline
Quantites &
Mean & SDDEV &  $1\sigma$-ranges & $2\sigma$-ranges & WMAP3 data ($\Lambda$CDM)  \\
\hline
$\alpha$ & $6.19$ & $3.31$ & $0.10-7.77$ & $0.10-11.82$ & ---       \\
$\beta$  & $0.42$ &$0.42$ & $< 0.47 $ & $< 1.36$ & ---       \\
$\Omega_B\, h^2[10^2]$ & $2.22$ & $0.07$ & $2.16 - 2.29$ & $2.09 - 2.35$ & $2.23\pm 0.07$  \\
$\Omega_{CDM}\, h^2[10^2]$ & $11.10$ & $0.65$ & $10.47-11.75$ & $9.85-12.40$ & $12.8\pm 0.8$ \\
$H_0$ & $65.37$ & $3.41$ & $61.84 - 68.73$ & $58.55 - 71.70$ & $72\pm 8$  \\
$Z_{re}$ & $10.89$ & $2.62$ & $4.00 - 12.26$ & $4.00 - 14.78$ &   ---     \\
$n_s$ & $0.95$ & $0.02$ & $0.94 - 0.97$ & $0.92 - 0.98$ & $0.958\pm 0.016$      \\
$A_s[10^{10}]$ & $20.69$ & $1.32$ & $19.31 - 22.00$ & $18.20 - 23.32$ & ---- \\
$\Omega_{Q}[10^2]$ & $67.90$ & $4.47$ & $63.74-72.16$ & $57.59-75.02$ & $71.6\pm 5.5$      \\
$Age/Gyrs$ & $13.97$ & $0.19$ & $13.78-14.16$ & $13.59-14.35$ &  $13.73\pm 0.16$ \\
$\Omega_{\nu}\,h^2[10^2]$ &$0.40$ & $0.32$ & $< 0.48$ & $< 0.91$ &
$< 1.93 (95\% C.L.)$ \\
$\tau$ & $0.084$ & $0.029$ & $0.053- 0.113$ & $0.028-0.139$ & $0.089
\pm 0.030$  \\ \hline
\end{tabular} \label{tab2}}
\TABLE[t]{
\caption{Global analysis data within $1\sigma$ and $2\sigma$
deviations for the exponential type (Model-III) of the quintessence
potential.} 
\begin{tabular}{@{}c||c|c||c|c||c@{}} \hline
Quantites &
Mean & SDDEV &  $1\sigma$-ranges & $2\sigma$-ranges & WMAP3 data ($\Lambda$CDM)  \\
\hline
$\alpha$ & $0.70$ & $0.42$ & $< 0.92$ & $< 1.41$ & ---       \\
$\beta$  & $0.50$ & $0.48$ & $< 0.58 $ & $< 1.53$ & ---       \\
$\Omega_B\, h^2[10^2]$ & $2.21$ & $0.07$ & $2.15 - 2.28$ & $2.08 - 2.34$ & $2.23\pm 0.07$  \\
$\Omega_{CDM}\, h^2[10^2]$ & $11.10$ & $0.63$ & $10.48-11.72$ & $9.84-12.33$ & $12.8\pm 0.8$ \\
$H_0$ & $65.61$ & $3.26$ & $62.37 - 68.70$ & $58.99 - 71.58$ & $72\pm 8$  \\
$Z_{re}$ & $11.07$ & $2.44$ & $10.07 - 12.35$ & $6.64 - 14.78$ &   ---     \\
$n_s$ & $0.95$ & $0.02$ & $0.94 - 0.97$ & $0.92 - 0.98$ & $0.958\pm 0.016$      \\
$A_s[10^{10}]$ & $20.73$ & $1.24$ & $19.48 - 21.95$ & $18.33 - 23.27$ & ---- \\
$\Omega_{Q}[10^2]$ & $68.22$ & $4.17$ & $64.38-72.08$ & $58.45-75.05$ & $71.6\pm 5.5$      \\
$Age/Gyrs$ & $13.96$ & $0.19$ & $13.77-14.15$ & $13.61-14.36$ &  $13.73\pm 0.16$ \\
$\Omega_{\nu}\,h^2[10^2]$ &$0.38$ & $0.25$ & $< 0.48$ & $< 0.84$ &
$< 1.93 (95\% C.L.)$ \\
$\tau$ & $0.086$ & $0.027$ & $0.058- 0.113$ & $0.032-0.140$ & $0.089
\pm 0.030$  \\ \hline
\end{tabular} \label{tab3}}

\section{Summary and conclusion:}
In summary, we investigated the  dynamics of dark energy in
mass-varying neutrinos. We showed and discussed many aspects
of the interacting dark-energy with neutrinos scenario: (1) To
explain the present cosmological observation data, we don't need to
tune the coupling parameters between neutrinos and quintessence
field, (2) Even with a inverse power law potential or exponential
type potential which seem to be ruled out from the observation of
$\omega$ value, we can receive that the apparent value of the
equation of states can pushed down less than -1, (3) As a
consequence of global fit, the cosmological neutrino mass bound
beyond $\Lambda CDM$ model was first obtained with the value 
$\sum m_{\nu} < 0.45 \, eV (68 \% C.L.) \,\,\, [0.87 \, eV (95 \% C.L.)]$.

\appendix
\section{Boltzman Equations in Interacting Dark Energy-Neutrinos Scenario}
From the Lagrangian ${\cal L} = - m(\phi) \sqrt{-g_{\mu\nu}
\dot{x_{\mu}} \dot{x_{\nu}} }$, the Euler-Lagrange equation is given
by
\begin{equation}
{d \over d\lambda}\left(\partial {\cal L} \over \partial
\dot{x}^{\mu} \right) = {\partial {\cal L} \over \partial x^{\mu}}
\label{eq:euler-lagrange}
\end{equation}
where
\begin{eqnarray}
{\partial{\cal L} \over \partial \dot{x^{\mu}}} &=&
 P_{\mu} = m(x^{\mu})\, g_{\mu\alpha} { \dot{x}^{\alpha} \over
 \sqrt{-g_{\mu\nu} \dot{x}^{\mu} \dot{x}^{\nu} } }, \\
{\partial{\cal L} \over \partial x^{\mu}} &=& - {\partial m \over
\partial x^{\mu}} \sqrt{-g_{\alpha\beta} \dot{x}^{\alpha}
\dot{x}^{\beta} }  + m(x^{\mu}) {g_{\alpha\beta,\mu}
\dot{x}^{\alpha} \dot{x}^{\beta}  \over 2 \sqrt{-g_{\alpha\beta}
\dot{x}^{\alpha} \dot{x}^{\beta} } }
\end{eqnarray} Therefore eq(\ref{eq:euler-lagrange}) becomes
\begin{equation}
{1 \over \sqrt{-g_{\alpha\beta} \dot{x}^{\alpha} \dot{x}^{\beta}}}
{d \over d \lambda} \left(m(x^{\mu})\, { \dot{x}^{\mu} \over
 \sqrt{-g_{\alpha\beta} \dot{x}^{\alpha} \dot{x}^{\beta} } } \right)
 + {m(x^{\mu}) \over 2} {g_{\alpha\beta,\mu} \dot{x}^{\alpha}
 \dot{x}^{\beta} \over g_{\alpha\beta} \dot{x}^{\alpha}
 \dot{x}^{\beta} } = - {\partial m \over \partial x^{\mu}}
 \label{eq:b3}
\end{equation}
By using the relation $ds = \sqrt{-g_{\alpha\beta} \dot{x}^{\mu}
\dot{x}^{\nu}} d\lambda$, we obtain
\begin{equation}
P^{\mu} = m(x^{\mu}) { \dot{x}^{\mu} \over \sqrt{-g_{\alpha\beta}
\dot{x}^{\alpha} \dot{x}^{\beta} } }  = m(x^{\mu}) {dx^{\mu} \over
ds}
\end{equation}
and eq.(\ref{eq:b3}) becomes
\begin{eqnarray}
{d \over ds} \left(m(x^{\mu}) g_{\mu\beta} {dx^{\beta}  \over ds}
\right) - {m(x^{\mu}) \over 2} g_{\alpha\beta,\mu} {dx^{\alpha}
\over ds}
{dx^{\beta} \over ds} &=& - {\partial m \over \partial x^{\mu}}, \\
\cr
 {d \over ds}(g_{\mu\beta} P^{\beta}) - {1 \over 2}
 g_{\alpha\beta,\mu} P^{\alpha} {dx^{\beta} \over ds} &=& - {\partial
 m \over \partial x^{\mu}}
\end{eqnarray}
With simple calculation, finally we obtain the relations:
\begin{eqnarray}
{d P^{\nu} \over ds} + \Gamma^{\nu}_{\alpha\beta} \, P^{\alpha} { d
x^{\beta} \over ds} &=& - g^{\nu\mu} {\partial m \over \partial
x^{\mu}} \\
P^0 {d P^{\nu} \over d\tau} + \Gamma^{\nu}_{\alpha\beta} \,
P^{\alpha} P^{\beta} &=& - m g^{\nu\mu} m_{,\nu}. \label{eq:b9}
\end{eqnarray}
For $\mu=0$ component, eq.(\ref{eq:b9}) can be expressed as
\begin{equation}
{1 \over 2} {d \over d\tau} (P^{0})^2 + \Gamma^0_{\alpha\beta} \, P^{\alpha} P^{\beta} =-m
g^{0\mu} m_{,\mu}. \label{eq:b10}
\end{equation}
Since $P^0 = g^{00} P_0= a^{-2} \epsilon$, each terms of the
eq.(\ref{eq:b10}) are given by:
\begin{eqnarray}
{\rm First \,\, term} &=& -2 a^{-4} H q^2 + a^{-4} q
{dq \over d\tau} - a^{-2} H m^2 + a^{-2} m {dm \over d\tau} \\
{\rm Second \,\, term} &=& 2 a^{-4} H q^2 + a^{-2} H m^{2} + a^{-4}
{1 \over 2} \dot{h}_{ij} q^{i} q^{j} \\
{\rm Third \,\, term} &=& a^{-2} m {\partial m \over \partial\tau}
\end{eqnarray}
Since the first term includes the total derivative w.r.t. comoving
time, we obtain finally the eq.(\ref{eq:eq-b}) in Section III-C:
\begin{equation}
{dq \over d\tau} = -{1 \over 2} \dot{h}_{ij}\, q \, n^{i} n^{j} -
a^2 {m \over q} {\partial m \over \partial x^{i}} \, {d x^{i} \over
d \tau}.
\end{equation}


\section{Varying Neutrino Mass in Early stage of Universe}
In the high-redshift region with $z > 100$, neutrinos behavior as
like relativistic particles, even though they have non-zero mass.
Since $1/\epsilon = [1-a^2 m^2/2\, q^2]/q$ we have
\begin{eqnarray}
(\rho-3 P_{\nu}) &=& a^{-4} \int {d^3 q \over (2\pi)^3} {a^2 m^2 \over \epsilon} \, f_0, \nonumber \\
&=& {a^{-4} \over (2\pi)^2} [\int q\, dq \, (a\,m)^2 \, f_0 -\int{dq
\over q} (a\, m)^4 \, f_0], \nonumber \\
&=& {a^{-4} \over (2\pi)^2} [{\pi^2 \over 12} (a\, m)^2 -{\cal O}
(a^4 \, m^4) ]
\end{eqnarray}
For the massless neutrino case, the energy density is govern by
\begin{equation}
(\rho_{\nu})_{massless} = a^{-4} \int{d^3 q \over (2\pi)^3} \,\, q
\,\, f_0 = {a^{-4} \over (2\pi)^2} {7 \over 120} \pi^4.
\end{equation}
Then when we normalize it w.r.t. the energy density of massless
neutrino:
\begin{equation}
{(\rho_{\nu} -3\, P_{\nu}) \over (\rho_{\nu})_{massless}} = {10
\over 7\, \pi^2} (a\, m)^2
\end{equation}
Now we solve the equation of motion of quintessence field from
eq.(\ref{eq:Qddot})with exponential type potential (Model III) for
simplicity:
\begin{eqnarray}
\ddot{\phi} + 2{\cal H} \dot{\phi} &=& -a^2 {dV \over d\phi} - a^2
{dV_I \over d\phi} \nonumber \\
&=& a^2 \alpha \, V(\phi) - a^2 {\beta \over M_{pl}} (\rho_{\nu} -3
\, P_{\nu})
\end{eqnarray}
In the relativistic neutrino case when they obey the slow-rolling
condition:
\begin{equation}
2{\cal H} \dot{\phi} \approx -a \,\, {\beta \over M_{pl}} \,\,
(\rho_{\nu} - 3\, P_{\nu}) \propto -\beta
\end{equation}
Since the relative sign of $\beta $ and $\phi$ is always opposite,
the slope of the varying neutrino mass become negative, this means
that the primordial neutrino mass was decreasing when universe was
expanded.

\section{Consistency check}
The form of $\kappa$ can be also obtained by demanding
conservations of energy and momentum, i.e., demanding that $\nabla_\mu
\st{$\phi$}{\delta T^\mu_{\nu}}+\nabla_\mu \st{$\nu$}{\delta
T^\mu_{\nu}}~=~0$.
Let us begin by considering the divergence of the perturbed
stress-energy tensor for the scalar field,
\begin{eqnarray}
 \nabla_\mu \st{$\phi$}{\delta T^\mu_{\nu}} &=&
  -a^{-2}\left(\ddot{\phi}+2{\cal
      H}\dot{\phi}+a^2\frac{dV}{d\phi}\right) \partial_\nu
  \delta\phi
-a^{-2}\left(\ddot{\delta\phi} +2{\cal
      H}\dot{\delta\phi}+k^2\delta\phi + a^2 \frac{d^2
      V}{d\phi^2}\right)\partial_\nu\phi\nonumber \\
&=&\delta\left(\frac{dV_I}{d\phi}\right)\partial_\nu{\phi}+\frac{dV_I}{d\phi}\partial_\nu{\delta\phi}
\label{eq:scalar_field_divergence}
\end{eqnarray}
where in the last line we used eqs.(\ref{eq:Qddot}) and (\ref{eq:dQddot}).
The divergence of the perturbed stress-energy tensor for the neutrinos
is given by,
\begin{equation}
\nabla_\mu \st{$\nu$}{\delta T^\mu_{0}} =
 -\dot{\delta\rho}-(\rho+P)\partial_i v_i-3{\cal H}(\delta\rho +\delta
 P)-\frac{1}{2}\dot{h}(\rho+P)
\label{eq:neutrino_time_divergence}
\end{equation}
for the time component and
\begin{equation}
\nabla_\mu \st{$\nu$}{\delta T^\mu_{i}} =
 (\rho+P)\dot{v_i}+(\dot{\rho}+\dot{P})v_i +4{\cal
 H}(\rho+P)v_i+\partial_i P +\partial_j \Sigma^j_i
\label{eq:neutrino_spatial_divergence}
\end{equation}
for the spatial component.
Let us check the energy flux conservation for example, starting with the
energy flux in neutrinos (in $k$-space):
\begin{equation}
(\rho_\nu+P_\nu)\theta_\nu=4\pi k a^{-4}\int q^2 dq qf_0(q)\Psi_1
\end{equation}
where $\theta_\nu = i k^i v_{\nu~i}$. Differentiate with respect to
$\tau$, we obtain,
\begin{eqnarray}
(\rho_\nu+P_\nu)\dot\theta_\nu+(\dot\rho_\nu+\dot P_\nu)\theta_\nu
&=& 4\pi k a^{-4}\int q^2dq q f_0 \dot \Psi_1
-4{\cal H} (\rho_\nu+P_\nu)\theta_\nu
\label{eq:theta_nu_dot}
\end{eqnarray}
Let us consider the first term in the right hand side of the above
equation. This gives
\begin{eqnarray}
4\pi k a^{-4}\int q^2dq q f_0 \dot \Psi_1&=& 4\pi k a^{-4} \int q^2 dq q
 f_0 \left[ \frac{1}{3}\frac{q}{\epsilon}k(\Psi_0-2\Psi_2)
      +\kappa\right] \nonumber \\
&=& k^2 \delta P_\nu -k^2(\rho_\nu +P_\nu)\sigma_\nu +\frac{1}{3} 4\pi
 k^2 a^{-4}\int q^2 dq \frac{q^2}{\epsilon^2}\frac{\partial
 \epsilon}{\partial \phi}\delta\phi f_0\nonumber \\
&&+ 4\pi k a^{-4}\int q^2 dq q f_0 \kappa \nonumber
\end{eqnarray}
where $\sigma$ is defined as $(\rho+P)\sigma=-(k_i k_j
-\frac{1}{3}\delta_{ij})\Sigma^i_j$ and expressed by the distribution
function as
\begin{equation}
(\rho_\nu+P_\nu)\sigma_\nu =\frac{8\pi}{3}a^{-4}\int q^2 dq \frac{q^2}{\epsilon}f_0(q)\Psi_2
\end{equation}
Comparing eq. (\ref{eq:theta_nu_dot}) with eq.(\ref{eq:neutrino_spatial_divergence}), we find that
the divergence of the perturbed stress-energy tensor in spatial part for
the neutrinos leads to
\begin{equation}
\partial^i \nabla_\mu \st{$\nu$}{\delta T^\mu_{i}} = \frac{1}{3} 4\pi
 k^2 a^{-4}\int q^2 dq \frac{q^2}{\epsilon^2}\frac{\partial
 \epsilon}{\partial \phi}\delta\phi f_0 +4\pi k a^{-4} \int
 q^2 dq q f_0 \kappa
\end{equation}
On the other hand, the divergence of the perturbed stress-energy tensor
in spatial part for scalar field is, from
eq.(\ref{eq:scalar_field_divergence}),
\begin{eqnarray}
\partial^i \nabla_\mu \st{$\phi$}{\delta T^\mu_i} &=& -k^2
 \delta\phi\left(\frac{\partial \ln m_\nu}{\partial
        \phi}\right)(\rho_\nu-3P_\nu)
 = -4\pi k^2 \delta\phi a^{-4} \int q^2 dq \frac{\partial
  \epsilon}{\partial \phi}f_0~.
\end{eqnarray}
These two equations imply that $\kappa$ shold take the form as
eq. (\ref{eq:kappa}).

Next let us check the energy conservation. Density perturbation in
neutrino is, (see eq.(\ref{eq:delta_rho_nu}))
\begin{equation}
\delta\rho_\nu=a^{-4}\int\frac{d^3 q}{(2\pi)^3} \epsilon
 f_0(q)\Psi_0+a^{-4}\int\frac{d^3 q}{(2\pi)^3}\frac{\partial
 \epsilon}{\partial \phi}\delta\phi f_0~,
\end{equation}
By differenciate with respect to $\tau$, we obtain
\begin{eqnarray}
\delta\dot\rho_\nu &=& -4{\cal H}\delta\rho_\nu + a^{-4}\int\frac{d^3
 q}{(2\pi)^3} \dot{\epsilon}f_0 \Psi_0+a^{-4}\int\frac{d^3 q}{(2\pi)^3}
 \epsilon f_0 \dot\Psi_0 \nonumber \\
&&+a^{-4}\int\frac{d^3 q}{(2\pi)^3} \frac{\partial}{\partial
 \tau}\left(\frac{\partial \epsilon}{\partial \phi}\right)\delta\phi
 f_0
+a^{-4}\int\frac{d^3 q}{(2\pi)^3}\frac{\partial \epsilon}{\partial \phi}\dot{\delta\phi}f_0
\end{eqnarray}
where
\begin{eqnarray}
\dot\epsilon &=& ({\cal H}a^2 m^2 +a^2 m^2 \frac{\partial \ln
m_\nu}{\partial \phi}\dot\phi)/\epsilon~, \nonumber \\
\frac{\partial}{\partial\tau}\left(\frac{\partial \epsilon}{\partial
                  \phi}\right) &=& -{\cal
H}\frac{a^2m^2}{\epsilon^2}\frac{\partial \epsilon}{\partial
\phi}+2{\cal H}\frac{\partial \epsilon}{\partial \phi}+\frac{\partial^2
\epsilon}{\partial \phi^2}\dot\phi
\end{eqnarray}
Inserting eq.(\ref{eq:dot_Psi_0})
for $\dot\Psi_0$ in the above equation, we obtain
\begin{eqnarray}
\delta\dot\rho_\nu &=& -3{\cal H}(\delta\rho_\nu+\delta
 P_\nu)-(\rho_\nu+P_\nu)\theta_\nu -\frac{1}{2}\dot{h}
 (\rho_\nu+P_\nu) \nonumber \\
&&+a^{-4}\int\frac{d^3 q}{(2\pi)^3} f_0 \left(\frac{\partial^2
                     \epsilon}{\partial
                     \phi^2}\delta\phi +\Psi_0
                     \frac{\partial
                     \epsilon}{\partial
                     \phi}\right)\dot\phi
+a^{-4}\int\frac{d^3 q}{(2\pi)^3}f_0 \frac{\partial \epsilon}{\partial \phi}\dot{\delta\phi}
\end{eqnarray}
Comparing with eq.(\ref{eq:neutrino_time_divergence}), we find
\begin{equation}
\nabla_\mu \st{$\nu$}{\delta T^\mu_{0}} =-a^{-4}\int\frac{d^3
 q}{(2\pi)^3} f_0 \left(\frac{\partial^2
                     \epsilon}{\partial
                     \phi^2}\delta\phi +\Psi_0
                     \frac{\partial
                     \epsilon}{\partial
                     \phi}\right)\dot\phi -a^{-4}\int\frac{d^3 q}{(2\pi)^3}f_0
                     \frac{\partial \epsilon}{\partial \phi}\dot{\delta\phi}~,
\end{equation}
which is found to be equal to $-\nabla_\mu \st{$\phi$}{\delta
T^\mu_{0}}=-\delta\left(\frac{d
V_I}{d\phi}\right)\dot\phi-\frac{dV_I}{d\phi}\delta\dot\phi$.




\vskip1.0cm
{\large \bf Acknowledgements:} \\
We would like to thank L. Amendola, S. Carroll, T. Kajino,
Lily Schrempp and O. Seto for useful comments and exciting discussions. 
K.I. thanks C.~van de Bruck for useful communications.
K.I.'s work is supported by Grant-in-Aid for JSPS Fellows. Y.Y.K's work
is partially supported by Grants-in-Aid for NSC in Taiwan, and Center 
for High Energy Physics(CHEP)/KNU and APCTP/Pohang in Korea.
K.I. thanks KICP and National Taiwan university for kind
hospitality where some parts of this work has been done. 
Y.Y. K. also thanks T. Kajino and NAOJ in Japan for kind
hospitality where some parts of this work has been done.



\begin{thebibliography}{999}
\bibitem{sn1a}
S.~Perlmutter et al.,Nature {\bf 391} (1998) 51[arXiv:astro-ph/7912212];
A. G.~Riess et al., Astrophys. J. {\bf 116} (1998) 1009[arXiv:astro-ph/980520];
S.~Perlmutter et al., ApJ {\bf 517} (1999) 565[arXiv:astro-ph/9812133].
\bibitem{wmap}
C.~L.~Bennett et al., Astrophys. J. Suppl. Ser. {\bf 148} (2003) 1;
J.~L.~Tonry et al.,Astrophys. J. {\bf 594} (2003) 1;
M.~Tegmark et al., Astrophys. J. {\bf 606} (2004) 702.
\bibitem{lambda}
L.~M.~Krauss and M.~S. Turner, Gen. Rel. Grav. {\bf 27} (1995) 1137 ;
P.~J.~E. Peebles and B.~Ratra, Reviews of Modern Physics, Vol{\bf 75} (2003) 559.

\bibitem{quintessence}
C.~Wetterich, Nucl. Phys. B{\bf 302} (1988) 645.

\bibitem{mgrav}
S.~M.~Carroll, M.~Trodden and M.~S.~Turner, Phys. Rev. D{\bf 70}:
043528 (2004).

\bibitem{seljak:0604335}
U. Seljak, A. Slosar, and P. McDonald, J. Cosmol Astropart. Phys. 10
(2006) 014.

\bibitem{Percival:2007yw}
  W.~J.~Percival, S.~Cole, D.~J.~Eisenstein, R.~C.~Nichol, J.~A.~Peacock, A.~C.~Pope and A.~S.~Szalay,
  Mon.\ Not.\ Roy.\ Astron.\ Soc.\  {\bf 381}, 1053 (2007)
  [arXiv:0705.3323 [astro-ph]].

\bibitem{Crittenden:1995ak}
  R.~G.~Crittenden and N.~Turok,
  Phys.\ Rev.\ Lett.\  {\bf 76}, 575 (1996)
  [arXiv:astro-ph/9510072].

\bibitem{Hu:2004yd}
  W.~Hu and R.~Scranton,
  Phys.\ Rev.\  D {\bf 70}, 123002 (2004)
  [arXiv:astro-ph/0408456].

\bibitem{Takada:2006xs}
  M.~Takada,
  Phys.\ Rev.\  D {\bf 74}, 043505 (2006)
  [arXiv:astro-ph/0606533].

\bibitem{Hannestad:2005ak}
  S.~Hannestad,
  Phys.\ Rev.\  D {\bf 71}, 103519 (2005)
  [arXiv:astro-ph/0504017].

\bibitem{Ichiki:2007vn}
  K.~Ichiki and T.~Takahashi,
  Phys.\ Rev.\  D {\bf 75}, 123002 (2007)
  [arXiv:astro-ph/0703549].


\bibitem{Carroll:1998zi}
  S.~M.~Carroll,
  Phys.\ Rev.\ Lett.\  {\bf 81}, 3067 (1998)
  [arXiv:astro-ph/9806099].

\bibitem{Bean:2000zm}
  R.~Bean and J.~Magueijo,
  Phys.\ Lett.\  B {\bf 517}, 177 (2001)
  [arXiv:astro-ph/0007199].

\bibitem{Farrar:2003uw}
  G.~R.~Farrar and P.~J.~E.~Peebles,
  Astrophys.\ J.\  {\bf 604}, 1 (2004)
  [arXiv:astro-ph/0307316].

\bibitem{das-khoury:2006}
S. Das, P. ~S. Corasaniti and J. Khoury, Phys. Rev. {\bf D 73},
083509 (2006).

\bibitem{Lee:2006za}
  S.~Lee, G.~C.~Liu and K.~W.~Ng,
  Phys.\ Rev.\  D {\bf 73}, 083516 (2006)
  [arXiv:astro-ph/0601333].

\bibitem{Liu:2006uh}
  G.~C.~Liu, S.~Lee and K.~W.~Ng,
  Phys.\ Rev.\ Lett.\  {\bf 97}, 161303 (2006)
  [arXiv:astro-ph/0606248].

\bibitem{Fardon:2003eh}
R.~Fardon, A.~E.~Nelson and N.~Weiner, JCAP 0410:005, 2004;
[arXiv:astro-ph/0309800].

\bibitem{mavanu}
D.~B.~Kaplan, A.~E.~Nelson and N.~Weiner, Phys. Rev. Lett. {\bf
93}:091801, (2004); R.~D.~Peccei, Phys. Rev. {\bf D71}:023527
(2005).

\bibitem{Afshordi:2005ym}
  N.~Afshordi, M.~Zaldarriaga and K.~Kohri,
  Phys.\ Rev.\  D {\bf 72}, 065024 (2005)
  [arXiv:astro-ph/0506663].

\bibitem{Bean:2007ny}
  R.~Bean, E.~E.~Flanagan and M.~Trodden,
  arXiv:0709.1128 [astro-ph].

\bibitem{Brookfield-b}
A.~W.~Brookfield, C.~van de Bruck, D.~F.~Mota, and D.~Tocchini-Valentini,
Phys. Rev. Lett , {\bf 96}: 061301,2006;
Phys. Rev. {\bf D73}:083515,2006.

\bibitem{Zhao:2006zf}
  G.~B.~Zhao, J.~Q.~Xia and X.~M.~Zhang,
  arXiv:astro-ph/0611227.

\bibitem{Brookfield:2005bz}
  A.~W.~Brookfield, C.~van de Bruck, D.~F.~Mota and D.~Tocchini-Valentini,
  Phys.\ Rev.\  D  {\bf 76}, 049901 (2007)


\bibitem{Dvali:2000hr}
  G.~R.~Dvali, G.~Gabadadze and M.~Porrati,
  Phys.\ Lett.\  B {\bf 485}, 208 (2000)
  [arXiv:hep-th/0005016].

\bibitem{Brane-World}
P. ~Bin\'{e}truy, C. ~Deffayet, U. ~Ellwanger and D. ~Langlois,
Phys. Lett. {\bf B 477} (2000) 285;
P. ~Bin\'{e}truy, C. ~Deffayet and D. ~Langlois,
Nucl. Phys. {\bf B 565} (2000) 269;
For a review of Brane World Cosmology, 
P.~Brax, C. van de ~Bruck and A. C. ~Davis, Rept. Prog. Phys. 67:2183-2232 (2004).

\bibitem{MOND}
J. D. ~Bekenstein, Phys. Rev. {\bf D70}:083509 (2004); 
Erratum-ibid. {\bf D71}:069901 (2005);
C. ~Skordis, D. F. ~Mota, P. G. Ferreira and C. ~Boehm,
Phys. Rev. Lett. {\bf 96}:011301 (2006).


\bibitem{review-DE}
E.~J. Copeland, M.~Sami, and S.~Tsujikawa, Int. J. Mod. Phys., {\bf
D15}: 1753 (2006).
\bibitem{COY}
T.~Chiba, T.~Okabe, M. Yamaguchi, Phys. Rev. {\bf D62}: 023511,
2000.
\bibitem{k-essence}
C.~A.~Picon, V.~F.Mukhanov, P.~J. Steinhardt, Phys. Rev. {\bf D63}:
103510, 2001.
\bibitem{phantom}
R.~R.~Caildwell, Phys. Lett. {\bf B 545}: 23, (2002).
\bibitem{quintom}
Z.~K. Guo,Y.-S. Piao, X.-M. Zhang and Y.-Z Zhang, Phys. Lett. {\bf B 608}, 177 (2005).


\bibitem{Bi:2003yr}
X.~J.~Bi, P.~h.~Gu, X.~l.~Wang and X.~M.~Zhang, Phys. Rev. {\bf
D69}:113007 (2004);
[arXiv:hep-ph/0311022].

\bibitem{Ma:1995ey}
C.~P.~Ma and E.~Bertschinger,
Astrophys.\ J.\  {\bf 455}, 7 (1995).






\bibitem{Anderson:1997un}
  G.~W.~Anderson and S.~M.~Carroll,
  arXiv:astro-ph/9711288.



\bibitem{Ratra:1987rm}
  B.~Ratra and P.~J.~E.~Peebles,
  Phys.\ Rev.\ D {\bf 37}, 3406 (1988).

\bibitem{Brax:1999gp}
  P.~Brax and J.~Martin,
  Phys.\ Lett.\ B {\bf 468}, 40 (1999)
  [arXiv:astro-ph/9905040].


\bibitem{Brax:2000yb}
  P.~Brax, J.~Martin and A.~Riazuelo,
  Phys.\ Rev.\ D {\bf 62}, 103505 (2000)
  [arXiv:astro-ph/0005428].

\cite{Copeland:1997et}
\bibitem{Copeland:1997et}
  E.~J.~Copeland, A.~R.~Liddle and D.~Wands,
  Phys.\ Rev.\ D {\bf 57}, 4686 (1998)
  [arXiv:gr-qc/9711068].



\bibitem{eos-SN1a:2004}
A.~G. Riess {\it et al.}, Astrophys. J. {\bf 607}, 665 (2004).

\bibitem{hoekstra-CFHT:2005}
H. Hoekstra et al. [CFHT collaboration], arXiv:astro-ph/0511089,
2005.

\bibitem{astier-snls:2006}
P. Astier et al. [SNLS collaboration], Astron. Astrophy. 447 (2006) 31[astro-ph/0510447].

\bibitem{spergel-wmap3:2006}
D. Spergel at al. [WMAP collaboration], Astrophys. J. Suppl. 170 (2007) 377 [astro-ph/0603449].


\bibitem{ratra-peebles:1988}
B. Ratra and P.~J.E. Peebles, Astrophys. J. Lett. {\bf 325} (1988)
17; Phys. Rev. {\bf D 37} 3406(1988).

\bibitem{wetterich:1988}
C. Wetterich, Nucl. Phys. {\bf B302} (1988) 645.

\bibitem{ferreira:1997}
P.~G Ferreira and M. Joyce, Phys. Rev. Lett{\bf 79} 4740 (1997)
\bibitem{brax-martin:1999-2000}
P. Brax and J. Martin, Phys. Lett. {\bf B468} (1999) 40.
\bibitem{frieman:1995}
J.~A. Frieman, C.~T. Hill, A.Stebbins and I. Waga, Phys. Rev.
Lett{\bf 75} 2077 (1995).

\bibitem{zlatev:1999}
I. Zlatev, L.-M. Wang and P.~J. Steinhardt, Phys. Rev. Lett{\bf 82}
896 (1999).

\bibitem{sahni-wang:2000}
V. Sahni and L.-M. Wang, Phys. Rev. {\bf D62} (2000) 103517.

\bibitem{sahni-starobinsky:2000}
V. Sahni and A.~A. Starobinsky, Int. J. Mod. Phys. {\bf D9} 373
(2000).

\bibitem{lopez-matos:2000}
L.~A. Urena-Lopez and T. Matos, Phys. Rev. {\bf D62} 081302 (2000).

\bibitem{albrecht:2000}
A. Albrecht and C. Skordis, Phys. Rev. Lett{\bf 84} 2076 (2000).

\bibitem{lee-olive-pospelov:2004}
S. Lee, K.~A. Olive and M. Pospelov, Phys. Rev. {\bf D70} 083503
(2004).

\bibitem{mainz}
J.Bonne et al., Nuch. Phys. B (Proc. Suppl.), 91 (2001) 273.

\bibitem{troitsk}
Ch. Weinheimer, Nucl. Phys. B (Proc. Suppl.), 118 (2003) 279.

\bibitem{heidelberg-moscow}
L. Baudis et al. [Heidelberg-Moscow Collaboration], Phys. Rev. Lett.
{\bf 83}, 41 (1999).

\bibitem{igex}
C.~E. Aalseth et al. [IGEX Collaboration], Phys. Rev. {\bf D 65},
092007 (2002); Phys. Rev. {\bf D 70}, 078302 (2004).


\bibitem{yyk-OMEG07}
Y.-Y. Keum, K. Ichiki and T. Kajino, "Neutrino Mass Bounds from the
$0\nu\beta\beta$ Decays and Large Scale Structures", Presented at
the 10th International Symposium on Origin of Matter and Evolution
of Galaxies, Dec. 4-7, 2007, Sapporo, Japan; arXiv:0803.2393;
Yong-Yeon Keum, K. Ichiki, T. Kajino and G. J. Mathews,
"Neutrino mass bounds from Neutrinoless Double Beta Decays and Cosmological Probes
including Ly-$\alpha$ data", in preparation. 

\bibitem{Magano:2005}
G.~Mangano et al., Nucl. Phys. {\bf B729} (2005) 221.
\bibitem{beacom:2004}
J.~F. Beacom, N.~F. Bell and S. Dodelson, Phys. Rev. Lett. {\bf 93}
121302 (2004).

\bibitem{dodelson:2006}
S. Dodelson, A. Melchiorri and A. Slasar, Phys. Rev. Lett. {\bf 97}
04031 (2006).

\bibitem{hannestad:2003}
S. Hannestad, JCAP {\bf 05} 004 (2003).

\bibitem{pierpaoli:2003}
E. Pierpaoli, Mon. Not. Roy. Astron. Soc. {\bf 342} L63 (2003).

\bibitem{hannestad:2005}
S. Hannestad, JCAP {\bf 0502} 011 (2005); arXiv astro-ph/0411475.

\bibitem{slosar:2006}
A Slosar, Phys. Rev. {\bf D73} 123501 (2006).

\bibitem{Hu:1997mj}
  W.~Hu, D.~J.~Eisenstein and M.~Tegmark,
  Phys.\ Rev.\ Lett.\  {\bf 80}, 5255 (1998)
  [arXiv:astro-ph/9712057].


\bibitem{silk:1980}
J.~R. Bond, G. Efstathiou and J. Silk, Phys. Rev. Lett. {\bf 45}
1980 (1980).

\bibitem{ichikawa:2005}
K. Ichikawa, M. Fukugita and M. Kawasaki, Phys. Rev. {\bf D71}
043001 (2005).


\bibitem{fukujita:2006}
M. Fukugita, K. Ichikawa, M. Kawasaki and O, Lahav, Phys. Rev. {\bf
D 74}, 027302 (2006).

\bibitem{elgaroy:2002}
O. Elgaroy et al., Phys. Rev. Lett{\bf 89} 061310 (2002); O. Elgaroy
and O. Lahav, JCAP 0304 (2003) 004.

\bibitem{sanchez:2005}
A.~G. Sanchez et al., Mon. Not. Roy. Astron. Soc. 366 (2006) 189.

\bibitem{hannestad:2004}
S. Hannestad, JCAP 0305 (2003) 004.

\bibitem{barger:2003}
V. Barger, D. Marfatia, and A. Tregre, Phys. Lett. {\bf B 595} 55
(2004).

\bibitem{spergel:2006}
D.~N. Spergel et al. [WMAP Collaboration], astro-ph/0603449.

\bibitem{goobar:2006}
A. Goodbar, S. Hannestad, E. Mortsell, and H. Tu, J. Cosmol.
Astropart. Phys. 06 (2006) 019.












\bibitem{Hinshaw:2006ia}
  G.~Hinshaw {\it et al.}(WMAP collaboration),
  arXiv:astro-ph/0603451.

\bibitem{Page:2006hz}
  L.~Page {\it et al.}(WMAP collaboration),
  arXiv:astro-ph/0603450.

\bibitem{Cole:2005sx}
  S.~Cole {\it et al.}  [The 2dFGRS Collaboration],
  Mon.\ Not.\ Roy.\ Astron.\ Soc.\  {\bf 362}, 505 (2005)
  [arXiv:astro-ph/0501174].


\bibitem{McDonald:1999dt}
  P.~McDonald, J.~Miralda-Escude, M.~Rauch, W.~L.~W.~Sargent, T.~A.~Barlow, R.~Cen and J.~P.~Ostriker,
  Astrophys.\ J.\  {\bf 543}, 1 (2000)
  [arXiv:astro-ph/9911196].

\bibitem{Croft:2000hs}
  R.~A.~C.~Croft {\it et al.},
  Astrophys.\ J.\  {\bf 581}, 20 (2002)
  [arXiv:astro-ph/0012324].


\bibitem{Goobar:2006xz}
  A.~Goobar, S.~Hannestad, E.~Mortsell and H.~Tu,
  JCAP {\bf 0606}, 019 (2006)
  [arXiv:astro-ph/0602155].

\bibitem{McDonald:2004xp}
  P.~McDonald, U.~Seljak, R.~Cen, P.~Bode and J.~P.~Ostriker,
  Mon.\ Not.\ Roy.\ Astron.\ Soc.\  {\bf 360}, 1471 (2005)
  [arXiv:astro-ph/0407378].


\bibitem{MCMC}
A.~Lewis and S.~Bridle,
 Phys. Rev. {\bf D66}, 103511 (2002).

\bibitem{WHEPP8:2004} 
Some useful comments of the numerical analysis code for the mass-varying neutrinos can be found in
S. ~Goswami et al., Pramana 63:1391-1406(2004);[arXiv:hep-ph/0409225].


\bibitem{Bjaelde:2007ki}
  O.~E.~Bjaelde, A.~W.~Brookfield, C.~van de Bruck, S.~Hannestad, D.~F.~Mota, L.~Schrempp and D.~Tocchini-Valentini,
  arXiv:0705.2018 [astro-ph].

\bibitem{Kaplinghat:2006jk}
  M.~Kaplinghat and A.~Rajaraman,
  Phys.\ Rev.\  D {\bf 75}, 103504 (2007)
  [arXiv:astro-ph/0601517].

\bibitem{Takahashi:2006jt}
  R.~Takahashi and M.~Tanimoto,
  JHEP {\bf 0605}, 021 (2006)
  [arXiv:astro-ph/0601119].




\end{thebibliography}
\end{document}